\documentclass[12pt, a4paper]{article} 
\usepackage{a4} 
\usepackage{amsmath,amssymb} 
\usepackage{graphicx} 
\usepackage{subfigure} 
\usepackage{color} 
\usepackage[bf,footnotesize]{caption} 
\usepackage{appendix} 
 
\graphicspath{{plots/}} 
\def \3{\ss }

\newcommand{\tr}{\operatorname{Tr}} 
\newcommand{\re}{\operatorname{Re}}

\newcommand{\beq}{\begin{equation}} 
\newcommand{\eeq}{\end{equation}} 
\newcommand{\beqn}{\begin{eqnarray}} 
\newcommand{\eeqn}{\end{eqnarray}}

\newcommand{\be}{\begin{equation}} 
\newcommand{\ee}{\end{equation}}

\def\ors{a}
\def\rmii{b}
\def\mns{c}
\def\val{d}
\def\nic{e}
\def\rmiii{f}
\def\liv{g}
\def\ect{h}
\def\hum{i}
\def\zur{j}

\begin{document} 
\begin{titlepage} 
  {\vspace{-0.5cm} \footnotesize
  \hfill \parbox{60mm}{DESY 08-022, IFIC/08-08,\\ 
                       FTUV-08-1302, MS-TP-08-3,\\ 
                       RM3-TH/08-5, ROM2F/2008/04,\\ 
                       SFB/CPP-08-17, LTH 762,\\
                       HU-EP-08/10}}\\[9mm] 
  \begin{center} 
    \begin{LARGE} 
      \textbf{Dynamical Twisted Mass Fermions with Light Quarks: \\ [2mm]
              Simulation and Analysis Details}  
    \end{LARGE} 
  \end{center} 
 
  \vskip 0.5cm 
  \begin{figure}[h] 
    \begin{center} 
      \includegraphics[draft=false]{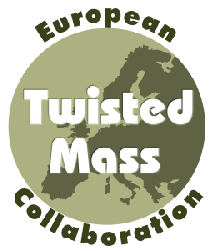} 
    \end{center} 
  \end{figure}

  \vspace{-0.8cm} 
  \baselineskip 20pt plus 2pt minus 2pt 

  \begin{center}
    \textbf{
      Ph.~Boucaud$^{(\ors)}$,
      P.~Dimopoulos$^{(\rmii)}$,
      F.~Farchioni$^{(\mns)}$,
      R.~Frezzotti$^{(\rmii)}$,
      V.~Gimenez$^{(\val)}$,
      G.~Herdoiza$^{(\nic)}$,
      K.~Jansen$^{(\nic)}$,
      V.~Lubicz$^{(\rmiii)}$,
      C.~Michael$^{(\liv)}$,
      G.~M\"unster$^{(\mns)}$,
      D.~Palao$^{(\val)}$,
      G.C.~Rossi$^{(\rmii)}$,
      L.~Scorzato$^{(\ect)}$,
      A.~Shindler$^{(\liv)}$,
      S.~Simula$^{(\rmiii)}$,
      T.~Sudmann$^{(\mns)}$,
      C.~Urbach$^{(\hum)}$,
      U.~Wenger$^{(\zur)}$}\\
  \end{center}
   
  \begin{center} 
    \begin{footnotesize} 
      \noindent  
 
      $^{(\ors)}$ Laboratoire de Physique Th\'eorique (B\^at.~210), Universit\'e
      de Paris XI,\\ Centre d'Orsay, 91405 Orsay-Cedex, France\\
      \vspace{0.2cm}

      $^{(\rmii)}$ Dip. di Fisica, Universit{\`a} di Roma Tor Vergata and INFN
      Sez. di Roma Tor Vergata,\\ Via della Ricerca Scientifica 1, I-00133 Roma, Italy\\
      \vspace{0.2cm}

      $^{(\mns)}$ Universit\"at M\"unster, Institut f\"ur Theoretische Physik,
      \\Wilhelm-Klemm-Strasse 9, D-48149 M\"unster, Germany\\
      \vspace{0.2cm}

      $^{(\val)}$ Dep. de F\'{\i}sica Te\`{o}rica and IFIC, Universitat de Val\`encia-CSIC,\\
      Dr.Moliner 50, E-46100 Burjassot, Spain\\
      \vspace{0.2cm}

      $^{(\nic)}$ DESY Zeuthen, Platanenallee 6, D-15738 Zeuthen, Germany\\
      \vspace{0.2cm}

      $^{(\rmiii)}$ Dip. di Fisica, Universit{\`a} di Roma Tre and INFN, Sez. di
      Roma III,\\ Via della Vasca Navale 84, I-00146 Roma, Italy\\
      \vspace{0.2cm}

      $^{(\liv)}$ Theoretical Physics Division, Dept. of Mathematical Sciences,
      \\University of Liverpool, Liverpool L69 7ZL, UK\\
      \vspace{0.2cm}

      $^{(\ect)}$ ECT* and INFN, strada delle tabarelle, 286 - 38100 Trento, Italy\\
      \vspace{0.2cm}

      $^{(\hum)}$ Humboldt-Universit{\"a}t zu Berlin, Institut f{\"u}r Physik,\\
      Newtonstr. 15, 12489 Berlin, Germany\\
      \vspace{0.2cm}

      $^{(\zur)}$ Institute for Theoretical Physics, ETH Z{\"u}rich, CH-8093 Z{\"u}rich,
      Switzerland\\

    \end{footnotesize} 
  \end{center} 
   
\end{titlepage} 
 
  \begin{abstract} 
    \noindent In a recent paper~\cite{Boucaud:2007uk} we presented   
    precise lattice QCD   
    results of our European Twisted Mass Collaboration (ETMC). They were  
    obtained by employing two mass-degenerate flavours of  
    twisted mass fermions at maximal twist. In the present paper we  
    give details on our simulations and the computation of physical observables.  
    In particular, we discuss the problem of tuning to maximal twist,  
    the techniques we have used to compute correlators and error  
    estimates. In addition, we provide more information on  
    the algorithm used, the autocorrelation times and scale determination, 
    the evaluation of disconnected contributions and  
    the description of our data by means of chiral perturbation theory formulae. 
  \end{abstract} 
 
\tableofcontents 
 
\section{Twisted mass fermions}
\label{sec:twismasfer} 
 
Dynamical Wilson twisted mass fermions, when tuned to maximal  
twist~\cite{Frezzotti:2000nk,Frezzotti:2003ni}, have been demonstrated  
to lead to precise results for mesonic quantities down to pseudo scalar 
masses $m_\mathrm{PS} \lesssim 300$~MeV. Results in the quenched case  
were discussed in refs.~\cite{Jansen:2003ir,Jansen:2005gf,Jansen:2005kk}  
and in the case of two mass-degenerate flavours of quarks in ref.~\cite{Boucaud:2007uk}. 
Preparatory simulations with twisted mass dynamical fermions were performed  
in~\cite{Farchioni:2004us,Farchioni:2004fs,Farchioni:2005tu,Farchioni:2005bh}. 
In ref.~\cite{Boucaud:2007uk} many of the details of our computations had to be  
omitted and it is the purpose of the present paper to supplement those          
and fill this gap. 
 
This paper is organized as follows. In this section we introduce 
twisted mass fermions and discuss the important issue of tuning to  
maximal twist. In section~\ref{sec:chargedmesons}, we give details about our techniques to  
compute charged correlators and in section~\ref{sec:neutralmesons} to compute  
neutral correlators and quark-disconnected contributions.  
In section~\ref{sec:algo} we discuss the algorithm details and explain 
our analysis techniques for obtaining reliable error estimates.  
In section~\ref{sec:scale_static_potential} we provide details of our 
computation of the force parameter $r_0$ and in section~\ref{sec:results} we give 
some results for the pseudoscalar mass and decay constant, the untwisted 
PCAC quark mass and the renormalization constant $Z_{\rm V}$. 
We use chiral perturbation theory to fit our data and we detail  
this procedure in section~\ref{sec:xpt}. We end with a short summary in  
section~\ref{sec:summary}.  
 
We begin with the Wilson twisted mass  
fermionic lattice action for two flavours of mass degenerate 
quarks, which reads 
(in the so called twisted basis~\cite{Frezzotti:2000nk,Frezzotti:2003ni} 
and with fermion fields having continuum dimensions) 
\begin{equation} 
  \label{eq:Sf} 
  \begin{split} 
    S_\mathrm{tm}^{\rm F} = &\, a^4\sum_x\Bigl\{ 
    \bar\chi_x\left[D_{\rm W}+ m_0 + i\gamma_5\tau_3\mu_q  
    \right]\chi_x\Bigr\}\, , \\ 
    & D_{\rm W} = \frac{1}{2}\gamma_\mu\left(\nabla_\mu+\nabla_\mu^{*}\right) 
    -\frac{ar}{2}\nabla_\mu\nabla_\mu^{*} \, ,
  \end{split} 
\end{equation} 
where $m_0$ is the bare untwisted quark mass and $\mu_q$ the bare twisted 
quark mass, $\tau_3$ is the third Pauli matrix acting in flavour space  
and $r$ is the Wilson parameter, which we set to $r=1$ in our simulations. 
We denote by $\nabla_\mu$ and $\nabla_\mu^{*}$ the gauge covariant nearest  
neighbour forward and backward lattice derivatives. 
The bare quark mass $m_0$ is related to the so-called hopping  
parameter $\kappa$, which we will often use in this paper, by   
$\kappa=1/(8+2am_0)$. Twisted mass fermions are said to be at  
{\em maximal twist} if the bare untwisted mass is tuned to its critical  
value, $m_\mathrm{crit}$. We will discuss later how this can be  
achieved in practice. 
 
In the gauge sector we use, for reasons explained in~\cite{Boucaud:2007uk},  
the so called tree-level Symanzik improved 
gauge action (tlSym)~\cite{Weisz:1982zw} which includes besides the 
plaquette term $U^{1\times1}_{x,\mu,\nu}$ also rectangular $(1\times2)$ Wilson loops 
$U^{1\times2}_{x,\mu,\nu}$. It reads  
\begin{equation} 
  \label{eq:Sg} 
    S_g =  \frac{\beta}{3}\sum_x\Biggl(  b_0\sum_{\substack{ 
      \mu,\nu=1\\1\leq\mu<\nu}}^4\{1-\re\tr(U^{1\times1}_{x,\mu,\nu})\}\Bigr.  
     \Bigl.+ 
    b_1\sum_{\substack{\mu,\nu=1\\\mu\neq\nu}}^4\{1 
    -\re\tr(U^{1\times2}_{x,\mu,\nu})\}\Biggr)\, , 
\end{equation} 
where $\beta$ is the bare inverse coupling and we set   
$b_1=-1/12$ (with $b_0=1-8b_1$ as dictated by the requirement 
of continuum limit normalization).  
Note that at $b_1=0$ this action becomes the usual Wilson plaquette gauge action. 
 
\subsection{Tuning to maximal twist} 
\label{sec:maximaltwist} 
 
One of the main virtues of Wilson twisted mass fermions 
is that by tuning the bare quark mass $m_0$ to its critical value  
an automatic $\mathrm{O}(a)$ improvement can be  
achieved such that expectation values of parity even operators scale to their  
continuum limit with $\mathrm{O}(a^2)$ discretization errors~\cite{Frezzotti:2003ni}. 
It was shown in the scaling test study carried out  
in~\cite{Jansen:2003ir,Jansen:2005gf,Jansen:2005kk} in the quenched case   
that $\mathrm{O}(a)$ improvement works extremely well for maximally 
twisted mass quarks. In this context, the method to tune to maximal  
twist by setting the so-called (untwisted) PCAC mass to zero 
(in the limit $\mu_q \to 0$) was found to be very successful, in agreement with 
theoretical considerations~\cite{Aoki:2004ta,Sharpe:2004ny,Frezzotti:2005gi}. 
In the present paper essentially the same approach to set to zero the (untwisted) PCAC mass  
\begin{equation} 
  \label{eq:mpcac} 
  m_\mathrm{PCAC} = \frac{\sum_\mathbf{x}\langle\partial_0 A_0^a(\mathbf x,t) 
    P^a(0)\rangle} {2\sum_\mathbf{x}\langle P^a(\mathbf x,t) P^a(0)\rangle}  \, , 
 \qquad a=1,2 \, ,
\end{equation} 
was followed, by evaluating~(\ref{eq:mpcac}) at large enough time separation, so that the pion  
ground state is dominant. {}For a definition of the (twisted basis) operators appearing in  
eq.~(\ref{eq:mpcac}) see eq.~(\ref{operators}) of Appendix~\ref{sec:appendixA}.  
 
In principle one could think of determining $am_\mathrm{crit}$ at each value 
of $a\mu_q$ at which simulations are carried out and then perform  
an extrapolation to vanishing $a\mu_q$ based on data satisfying the bound     
$a\mu_q\gtrsim a^3\Lambda_\mathrm{QCD}^3$~\cite{Frezzotti:2005gi}. 
This method is, however, rather CPU-time expensive. 
We therefore prefer to determine the value of $am_\mathrm{crit}$ (at each fixed value of $\beta$)  
from the simulation at the lowest available value, 
$a\mu_{q,\mathrm{min}}\ll a\Lambda_\mathrm{QCD}$. This choice simply affects 
the critical quark mass by $\mathrm{O}(a\mu_{q,\mathrm{min}}\Lambda_\mathrm{QCD})$ terms. 
Therefore $\mathrm{O}(a)$ improvement is still guaranteed~\cite{Frezzotti:2003ni}.  
{}Furthermore, and most importantly, with such a determination of $am_\mathrm{crit}$ 
also the $\mathrm{O}(a^2)$ cutoff effects remain small as long as 
$\mu_q \gtrsim a^2\Lambda_{QCD}^3$~\cite{Frezzotti:2005gi}. 
We recall below the line of arguments leading to this conclusion.  

\subsection{Maximal twist and residual $\mathrm{O}(a^2)$ artifacts}
\label{subsec:1_2} 

To start the discussion let us assume that $m_0 = 1/(2\kappa) -4$ has been set 
to a value corresponding to some sensible lattice estimate of the critical mass, 
while $\mu_q$ is non-zero. In this situation one is already at maximal twist. 
However the unavoidable $\mathrm{O}(a)$ terms affecting any determination of the critical 
mass can be further tuned in an ``optimal way'', i.e.\ in a way such that the 
residual $\mathrm{O}(a^2)$ lattice artifacts in physical quantities remain under control 
as the pion mass is decreased. We briefly explain how this can be achieved  
in practice and to what accuracy, following the work of ref.~\cite{Frezzotti:2005gi}. 
In the Symanzik expansion of the lattice expectation value  
$\langle O(x)\rangle|_{\mu_q}$ of a multilocal operator $O$ computed  
at a bare quark mass $\mu_q$ there will appear at $\mathrm{O}(a^2)$  
terms which are proportional to 
\begin{equation} 
  \label{eq:xiovmpi} 
  \left[ \frac{1}{m_\pi^2}\right]^2\xi_\pi^2(\mu_q)\ \propto\ 
    \frac{1}{\mu_q^2}\xi_\pi^2(\mu_q) \, , 
\end{equation} 
where 
\begin{equation} 
  \label{eq:xipi} 
  \xi_\pi(\mu_q) = |\langle\Omega|\mathcal{L}_\mathrm{odd}|\pi^0(\mathbf{0}) 
  \rangle  |^\mathrm{cont}_{\mu_q}\, . 
\end{equation} 
Here $\langle\Omega |$ and $|\pi^0(\mathbf{0})\rangle$ denote the vacuum and 
the one-pion neutral state at zero three-momentum, respectively. With the symbol  
\begin{equation} 
  \label{eq:lodd} 
\mathcal{L}_\mathrm{odd}= a \mathcal{L}_5+a^3 \mathcal{L}_7+... 
\end{equation} 
we indicate the set of operators of odd dimension in the Symanzik 
local effective Lagrangian that describes  
the maximally twisted lattice theory. {}From eq.~(\ref{eq:xiovmpi}) one recognizes  
that cut-off effects may become large when $m_\pi^2$ gets small. 
 
The general strategy to avoid these large cut-off effects is to 
tune $\xi_\pi(\mu_q)$ to zero, or at least to reduce it to  
$\mathrm{O}(am_\pi^2\Lambda_\mathrm{QCD}^2)$ by adjusting the 
value of $\kappa_\mathrm{crit}$. One way to realise this is precisely to tune 
$m_\mathrm{PCAC}=0$ as explained above. In particular it is sufficient to 
impose the vanishing of the PCAC mass at $\mu_q=\mu_{q,\mathrm{min}}$~\cite{Frezzotti:2005gi}.  
An analysis \`a la Symanzik of the correlator in the numerator of eq.(\ref{eq:mpcac}) 
shows that, if $\kappa$ is such that $m_\mathrm{PCAC}$ vanishes at a given value 
of $\mu_q$ (provided $\mu_q <\Lambda_\mathrm{QCD}$), then $\xi_\pi(\mu_q)$ is suppressed 
in a sufficiently strong way, namely one gets (note that $\xi_\pi$ has mass dimension three) 
\begin{equation} 
  \label{eq:xieq} 
  \xi_\pi(\mu_q) = \mathrm{O}(a\mu_q\Lambda_\mathrm{QCD}^3) + 
  \mathrm{O}(a\mu_q^2\Lambda_\mathrm{QCD}^2) + 
  \mathrm{O}(a^3\Lambda_\mathrm{QCD}^6)\, ,
\end{equation}  
and thus (see eq.~(\ref{eq:xiovmpi}))
\begin{equation} 
  \label{eq:ximu} 
  \frac{\xi_\pi(\mu_q)}{\mu_q\Lambda_\mathrm{QCD}^2} = 
  \mathrm{O}(a\Lambda_\mathrm{QCD}) + \mathrm{O}(a\mu_q) + 
  \mathrm{O}(a\Lambda_\mathrm{QCD}\frac{a^2\Lambda_\mathrm{QCD}^3}{\mu_q}) \, . 
\end{equation} 
In this situation, the ratio $\xi_\pi(\mu_q)/\mu_q\Lambda_\mathrm{QCD}^2$  
remains small as long as $\mu_q \gtrsim a^2\Lambda_\mathrm{QCD}^3$. 
 
{}For each value of $\mu_q$ in the region $a^2\Lambda_{QCD}^3\lesssim\mu_q < \Lambda_\mathrm{QCD}\ll a^{-1}$ 
the value of $\kappa$ at which $m_\mathrm{PCAC}$ vanishes provides a legitimate estimate of  
$\kappa_\mathrm{crit}$ and hence of $m_\mathrm{crit}$.  
Estimates of $m_\mathrm{crit}$ corresponding to different values 
of $\mu_q$ differ by $\mathrm{O}(a\mu_q\Lambda_{\rm QCD})$ from each other. 
In particular, working at $\kappa_\mathrm{crit}(\mu_{q,{\rm min}})$ leads to  
$\mathrm{O}(a^2)$ cutoff effects which are at worst of the form  
$a^2(\mu_{q,{\rm min}}/\mu_q)^2$ and thus perfectly tolerable as long as  
$\mu_q \gtrsim \mu_{q,{\rm min}} > a^2\Lambda_{\rm QCD}^3$. 
This result can be checked by expanding $\xi_\pi(\mu_q)$ around  
$\mu_{q,{\rm min}}$ in eq.~(\ref{eq:ximu}) and using the expression of  
$\xi_\pi(\mu_{q,{\rm min}})$ from eq.~(\ref{eq:xieq}).   
 
\subsection{Numerical precision for tuning to maximal twist}
\label{subsec:1_3}

It remains to be discussed to what numerical precision the condition 
$m_\mathrm{PCAC}=0$ has to be fulfilled. This question is important if 
one wants to avoid that numerical uncertainties jeopardize the tuning procedure.  
Suppose $|am_\mathrm{PCAC}| = a\epsilon \neq 0$, where $a\epsilon$ denotes  
a small deviation, due to numerical limitations, from the condition of vanishing 
PCAC mass. As a rule of thumb the value of $a\epsilon$ can be taken as 
the maximum (in modulus) between the finite statistics central value  
of $am_\mathrm{PCAC}$ and its (estimated) standard deviation. It then follows by  
expanding $\xi_\pi = \xi_\pi(\mu_q,\epsilon)$ around $\epsilon=0$ 
\begin{equation} 
  \label{eq:xidelta} 
  \begin{split} 
    \xi_\pi(\mu_q,\epsilon) &= \xi_\pi(\mu_q) + \mathrm{O}\left( 
      \Lambda_\mathrm{QCD}^2 \epsilon\right) \\ 
    & \approx \mathrm{O}(a\mu_q\Lambda_\mathrm{QCD}^3) + 
    \mathrm{O}(a\mu_q^2\Lambda_\mathrm{QCD}^2) + 
    \mathrm{O}(\Lambda_\mathrm{QCD}^2 \epsilon)\, . \\ 
  \end{split} 
\end{equation} 
Thus for the relative size of $\xi_\pi$ compared to the 
actual value of the quark mass, one gets  
\begin{equation} 
  \label{eq:relprec} 
  \frac{\xi_\pi(\mu_q,\epsilon)}{\mu_q\Lambda_\mathrm{QCD}^2} = 
  \mathrm{O}(a\Lambda_\mathrm{QCD}) + \mathrm{O}(a\mu_q) + \mathrm{O}(\frac{\epsilon}{\mu_q}) \, . 
\end{equation} 
A smooth approach to the continuum is, of course, guaranteed when  
$|\epsilon/\mu_q |$ is  
of order $a\Lambda_\mathrm{QCD}$ or smaller. 
In fact, from the form of the dimension five term in 
the Symanzik effective Lagrangian of the twisted mass lattice QCD, 
it follows that, close to maximal twist, $a\Lambda_\mathrm{QCD} |\epsilon/\mu_q |$ 
is the expected order of magnitude of the (unwanted) relative $\mathrm{O}(a)$ cutoff effects  
stemming from violations of the condition of vanishing  
PCAC mass. The requirement 
$|\epsilon/\mu_q | \lesssim a\Lambda_\mathrm{QCD}$ thus implies that the  
relative magnitude of these unwanted cutoff effects satisfies the constraint 
$a\Lambda_\mathrm{QCD} |\epsilon/\mu_q | \lesssim a^2\Lambda_\mathrm{QCD}^2$, 
which is consistent with $\mathrm{O}(a)$ improvement.  
 
In practice, since we are interested in 
simulations performed at lattice spacings about (or slightly below) 0.1~fm, 
where $a\Lambda_\mathrm{QCD} \sim 0.1$, a value of $|\epsilon/\mu_q| \lesssim 0.1$  
(and ideally decreasing with $a$) will represent an acceptable 
level of precision in the procedure of determining the critical mass.  
This condition is not too restrictive as in actual simulations 
it is sufficient that it holds at $\mu_{q,{\rm min}}$. 
We also remark that in order  
to check scaling and perform a reliable continuum extrapolation, the value of  
$\mu_{q,{\mathrm min}}$ should be kept roughly fixed in physical units as the  
lattice spacing is decreased. 

Although these theoretical arguments show that we  
can work in conditions such that we are effectively left with 
only $\mathrm{O}(a^2)$ lattice artefacts, numerical 
computations are required to check the scaling behaviour and  
determine the order of magnitude of the coefficient multiplying 
$a^2$ terms for the observables of interest. In this paper, 
where data at only one value of $a$ are analyzed,  
we cannot evaluate these coefficients. Nevertheless, for the observables 
we discuss here preliminary results from our collaboration presented  
in ref.~\cite{Urbach:2007rt,Dimopoulos:2007qy} indicate that 
the residual cutoff effects are indeed small and  
consistent with $\mathrm{O}(a)$ improvement.

\section{Computations in the charged meson sector}
\label{sec:chargedmesons}

In this paper we will be mainly using the twisted quark basis where the
fermionic action takes the form~(\ref{eq:Sf}). Even though there is no
fundamental reason for this choice, employing the twisted quark basis 
makes immediately transparent the way several computational methods, 
which have been invented for, or widely applied to, untwisted Wilson fermions,
carry over to the case of maximally twisted Wilson quarks. Of course, in
such an unphysical basis, the two flavour components of the fermion field
$\chi = (u,d)^T$ appearing in the action do not coincide with the canonical
quark fields in the ``physical'' basis, $\psi = (u_{\rm phys},d_{\rm phys})^T$,
rather the former are related to the latter by the axial rotation
\begin{equation}
\chi = e^{-i\gamma_5\tau_3\omega/2}|_{\omega =\pi/2} \psi  \qquad
\Leftrightarrow \qquad u = e^{-i\gamma_5\pi/4} u_{\rm phys} \, , \quad
                d = e^{i\gamma_5\pi/4} d_{\rm phys} \, ,
\label{eq:tm2ph_quark}
\end{equation}
 which we write here in the case of maximal twist, $\omega=\pi/2$. Since
the axial transformation above is flavour diagonal, the names of the
components ($u$,$d$) of the twisted basis field $\chi$ are still
appropriate to their flavour content. In spite of that, the correct
interpretation of gauge invariant composite bare operators in the
($\chi$, $\bar\chi$) basis is obtained  only once they are expressed in
terms of the physical basis bare fields ($\psi$, $\bar\psi$).  Examples
concerning quark bilinear fields can be found in
Appendix~\ref{sec:appendixA}. 

In this context it may be useful to remark that, since parity and
isospin are no longer exact  symmetries (recall however that $I_3$, the
third isospin component, is unbroken), a physical  basis bare composite
operator with given formal parity and isospin properties can interpolate
 a hadron with opposite parity and/or different isospin. As a
consequence in the quantum-mechanical representation of the correlators
there will be contributions containing matrix elements of a physical
basis composite operator with given formal parity and isospin
between the vacuum and a state with opposite parity and/or different
isospin, as well as between a neutral pion state (which has the same
lattice quantum numbers as the vacuum) and a state with the same parity
and isospin properties as the considered operator. Such parity- and/or
isospin-violating matrix elements are of course of order $a$. Their
occurrence in the quantum-mechanical representation of correlators is 
not in contradiction with the ${\mathrm{O}}(a)$ improvement of the expectation 
values of parity-even, or isospin-invariant, multilocal
operators~\cite{Frezzotti:2003ni}.  For these specific correlators,
indeed, an analysis \`a la Symanzik  shows that each term of their
quantum-mechanical representation can contain only an even number of
${\mathrm{O}}(a)$ factors given by parity- and/or isospin-violating matrix
elements~\footnote {This result essentially follows from the property
that, at maximal twist, the order $a$ piece of the Symanzik effective
Lagrangian, $a\mathcal{L}_5$, is odd under parity and the flavour
exchange $u \leftrightarrow d$.}.

From the formulae in Appendix~\ref{sec:appendixA}, it is clear that at
maximal twist, $\omega=\pi/2$, the operator $\bar{d} \gamma_5 u$ is 
associated to the $\pi^+$ meson, in the sense that $(\bar{d} \gamma_5
u)^\dagger$ creates the $\pi^+$ state from the vacuum. The two-point
$\pi^+$ meson correlator receives contributions only from (fermionically) 
connected diagrams, and after integration over fermion fields, it is given by
 \begin{equation}
  C(t)= \langle {\rm tr} [ G_u(0,t) \gamma_5 G_d(t,0) \gamma_5 ] \rangle \, ,
 \label{eq:C2pi+}
 \end{equation}
where $\langle\ldots\rangle$ means average over the gauge ensemble,
the trace ${\rm tr}[\dots]$ is restricted to spin and colour indices only,
and
we denote by $G_u(0,t)$ the propagator for a $u$-quark from 0 to $t$, 
and correspondingly by $G_d$ the similar propagator for the $d$-quark. 
Here three-space indices are understood as at this stage we need not specify the
spatial separation, or equivalently the three-momentum.
We can use the identity~\footnote{Here (with a little abuse of notation)
by $^{+}$ we mean complex conjugation and transposition with respect to
spin-colour indices only, while $y = ({\bf y},y_0)$
and $z = ({\bf z},z_0)$ are the spacetime coordinates.} 
$G_d(y,z)=\gamma_5 G_u(z,y)^+ \gamma_5$ to
relate the connected correlator~(\ref{eq:C2pi+}) to propagators from a common
source (at time $x_0=0$) through 
\begin{equation}
  C(t)= \langle {\rm tr} [ G_u(0,t)G_u(0,t)^+ ] \rangle\, .
\end{equation}
Thus only propagators for one flavour at one source point  are needed
for the computation of the charged meson correlator.  As we discuss
later, it is more efficient, however, to evaluate correlation functions 
from a wider set of sources. 

In the Table below we give the correspondence between bilinear operators
of the form $\bar{d} \Gamma u$, where $\Gamma$ is an hermitian
combination of Dirac $\gamma$-matrices, and the mesonic state that is
associated with each of them (in the limit $a \to 0$, i.e.\ neglecting
${\mathrm{O}}(a)$ contamination from states of different parity and isospin). 

\begin{center}
\begin{tabular}{|c|c|} 
\hline
Meson & Operator \\
\hline
                    &     \\
$\pi^\pm$,\ $\pi^\pm$,\ $X_1^\pm$ &
  $\bar{d} \gamma_5 u$, $\bar{d} \gamma_0 u$, $\bar{d} i\gamma_0 \gamma_5  u$ \\
$\rho^\pm$,\ $\rho^\pm$,\ $a_1^\pm$ &
  $\bar{d} i\gamma_i \gamma_0  u$, $\bar{d}i\gamma_i \gamma_5  u$, $\bar{d} \gamma_i u$ \\
$b_1^\pm$ & $\bar{d} i\gamma_i \gamma_0 \gamma_5   u$ \\
$a_0^\pm$ & $\bar{d}    u$ \\
\hline
\end{tabular}
\end{center}
 In this Table, $X_1^\pm$ labels an isotriplet state with
$J^{P}=0^{+}$, for which there is no experimental candidate. We note
that the associated operator is in the continuum a component of a
conserved current in the theory with two mass degenerate quarks.

 We evaluate the two-point (connected) correlators for all the pairs of
operators in the same line of the Table above. In view of the symmetries
of the  lattice theory at maximal
twist~\cite{Frezzotti:2000nk,Frezzotti:2003ni},  such correlators are in
general non zero: e.g.\ the correlator obtained from the insertion of
the first (or second) operator in the second line of the Table with the
third operator in the same line is an ${\mathrm{O}}(a)$ quantity (in fact
$\rho^\pm$ and $a_1^\pm$ carry different continuum quantum numbers). 
Since we also use a local and extended (fuzzed) source and sink in all
cases we consider, we will have either $6 \times 6$ or  $2 \times 2$
matrices of correlators available.

Therefore, we measure in general correlation functions of several
different pairs of operators ($\langle O_{\alpha} O_\beta \rangle$,
with $\alpha,\beta=1,\dots N$) at source and
sink. We then use a factorizing fit expression where
$i=1,\dots M$ states 
(with energy denoted by $E_i$) are included
\begin{equation}
  C_{\alpha \beta}(t)= \sum_{i=1}^M c^{i}_{\alpha} c^{i}_{\beta}
 ( e^{-E_i t} \pm e^{-E_i (T-t) } )\, .
\label{eq:factorisingfit}
 \end{equation}
Here $T$ is the lattice temporal extent and the $\pm$ sign is determined by
the properties of the chosen operators under time-reflection.
 By simultaneously fitting  $N \times N$ correlators with $M$ 
 states, we can optimally determine energies and couplings. From them we
evaluate other quantities of interest, such as $af_{\pi}$ and
$am_{\mathrm{PCAC}}$. 
 We use  conventional methods to determine the  optimal  $t$ range, $N$-
and $M$-values to be employed in the fits. We take into account
statistical correlations among observables~\cite{Michael:1994sz} through
correlated fits to establish that the $\chi^2$ value  is acceptable. 
 Our final  fitted values are obtained from uncorrelated fits,
since that introduces less bias~\cite{Michael:1994sz}, although 
the $\chi^2$ values are smaller than those obtained including correlations.
 We also checked that the fits are stable when taking into account
correlations.
 For pseudoscalar mesons we use mainly $M=1$ as well as $N=4$ or $6$,
and select the minimum value of $t$  such that the effective masses (or
energies)  from different matrix elements agree. 

We conclude by recalling that, owing to reflection invariance of the
lattice action (the Euclidean analog of the Minkowski complex-conjugation, 
a symmetry that is preserved by Wilson fermions, either chirally twisted or not, see 
ref.~\cite{Frezzotti:2003ni}), all the correlators 
that are expectation values of fields with definite reflection properties 
are either real or purely imaginary, depending on whether the whole field product 
has even or odd reflection-parity. In particular, the expectation values 
of multilocal fields with negative spatial parity, which are ${\mathrm{O}}(a)$ 
quantities, come out to be purely imaginary if one does not take care
of inserting the $i$-factors that are needed to render the multilocal field 
even (rather than odd) under the reflection. 

\subsection{Quark propagators from stochastic sources}
\label{subsec:sources}

Although it is feasible to use $u$-quark propagators from 12 colour-spin
 sources (with each source being non-zero only for one colour-spin
combination) at one space-time point to evaluate mesonic correlators, it
is preferable to use the information contained in the gauge
configurations more fully, especially in the case of such CPU-time
expensive simulations as dynamical quark simulations.  One efficient way
to achieve this goal is to use  stochastic sources. To keep the
noise-to-signal ratio reasonable, it is  mandatory to use time-slice
sources rather than full volume sources. A great reduction of the
noise-to-signal ratio over conventional stochastic methods (see
ref.~\cite{Foley:2005ac} for a review)  can be
obtained~\cite{Foster:1998vw,McNeile:2006bz} by using the
``one-end-trick'' which is described below. A similar method, 
called random wall, was used by MILC~\cite{Aubin:2004fs}. 

The starting point of all stochastic methods for computing quark propagators
is the introduction of random sources, $\xi^{r}_i$, where $i=1,\dots V_s$ spans 
the set of the source degrees of freedom (colour, spin, space, time) and $r=1,
\dots R$ labels the noise samples generated for each gauge configuration. The
corresponding average satisfies
 \begin{equation}
\lim_{R \to \infty} [\xi_i^* \xi_j]_R = \delta_{ij},\ \  
\lim_{R \to \infty} [\xi_i \xi_j]_R=0 \, , 
\label{oenend1}
 \end{equation}
which can be achieved by various different noise choices, such as  
$\xi^{r}_i=(\pm 1 \pm i)/\sqrt{2}$ or gaussian (complex) noise. 

As a next step, we invert the lattice Dirac matrix $M$ (for one given 
quark flavour) on each sample of this source,
\begin{equation}
  \phi^{ r}_j=M^{-1}_{jk} \xi^{  r}_k \, ,
  \label{oneend2}
\end{equation}
so that averaging over $r$ ({$R$} samples) gives
\begin{equation}
  [\xi^{{  r}*}_i \phi^{  r}_j ]_R =
  [\xi^{{  r}*}_i M^{-1}_{jk} \xi^{  r}_k ]_R
  =  M^{-1}_{ji} + {\rm noise} \, ,
  \label{oneend3}
\end{equation}
 where $j$ can be arbitrary and $i$ belongs to the set of indices for
which the source is non-vanishing, which we assume to be of size $V_s$.
 The quantity~(\ref{oneend3}) is an unbiased estimator of the quark
propagator from $i$ to $j$. Unfortunately, here the noise is expected 
to be as $\approx \sqrt{V_s}/\sqrt{R}$ whereas the signal is $\approx 1$ 
at best. Variance reduction is thus very necessary. Furthermore for a meson 
correlator, the signal behaves as $\exp(-m_{\rm meson}t)$ which decreases 
rapidly with increasing $t$.

The `one-end-trick' allows~\cite{Foster:1998vw,McNeile:2006bz} a more
favourable signal-to-noise ratio.  Consider the product $\phi^{r*}_i
\phi^{r}_j$ where the stochastic source is now non-zero for all
colour-spin indices and all space points at only one time, denoted by
$t_0$  (time-slice source).  Then upon averaging  over $r$ one has 
 \begin{equation}
  [\, \phi^{r*}_i \phi^{r}_j \, ]_R =
  [\, (M^{-1}_{ik} \xi^{r}_k)^*
  M^{-1}_{jm} \xi^{  r}_m \, ]_R  =  M^{-1*}_{ik} M^{-1}_{jk}
  + {\rm noise} \, ,
  \label{onened4}
 \end{equation} 
 where the sum over $k$  includes all source components. This quantity
is an unbiased estimator for the  
product of the quark propagators $M^{-1}_{jk} M^{-1\;+}_{ki}$ 
from the source to sites $i$ and $j$ on each gauge configuration. 
Then contracting with $\delta_{ij}$ and summing over 
space at fixed time-slice $t$  yields the full zero three-momentum
($\pi^\pm$-channel) correlator from $t_0$ to $t$. The noise counting is
now more favourable. There are $V_s^2$ noise terms, 
which yield a standard deviation of order $V_s$, 
but the signal itself is of order $V_s$. This is such
big an advantage that it is sufficient to employ just one sample of
noise per gauge configuration ($R=1$). As we discuss below, the optimal
way to choose the time-slice ($t_0$) at which  the stochastic source is
located, is to change it randomly as the gauge configuration is changed.
It should be remarked that the `one-end-trick', as formulated 
above, only works for the case of a zero three-momentum interpolating
field of the form $\bar{d} \gamma_5 u$ at the source time ($t_0$).

A convenient extension of the `one-end-trick', that allows meson-to-meson
correlators with any Dirac structure at the source to be evaluated, 
requires consideration of four ($\beta=1,2,3,4$) ``linked'' sources of the
form 
 \[ \xi_{\alpha,c,{\bf x},x_0}^{(\beta ; \; t_0)} = \delta_{\alpha\beta}
\delta_{x_0 t_0} \eta_{c,{\bf x}} \, , \]
where $\alpha$ and $c$ are
Dirac and colour indices respectively, while $\eta$ is a non-vanishing 
noise field. Such sources, which are non-zero only on a given time-slice
($t_0$) and when the Dirac index value equals $\beta$, are called
``linked'' because they involve a common noise field $\eta$.\,\footnote {Note that ``linked'' sources are different than ``spin-diluted'' sources~\cite{Foley:2005ac,Gusken:1999as} since these require different random numbers for each spin.} One can
check that by replacing $\xi$ and $\xi^*$ in the l.h.s.\ of
eq.~(\ref{onened4}) by two of these linked sources, say $\xi^{(\beta ; \; t_0)}$
and $\xi^{(\gamma ; \; t_0)\;*}$, and choosing appropriately $\beta$ and $\gamma$,
it is possible to evaluate the two-point correlators with a field of the
form $\bar{d} \Gamma u$ at the source ($x_0=t_0$) with any Dirac matrix
$\Gamma$. This very useful extension, which we have thoroughly exploited
in the present paper, comes at a moderate price. One must in fact only
perform four separate inversions (per gauge configuration and per noise 
sample), one for each of the four linked sources $\xi^{(\beta ; \; t_0)}$,
$\beta=1,\dots 4$.

To further extend the one-end trick with linked sources to non-zero
three-momentum or to spatially  non-local mesonic operators is
completely straightforward, at the cost of more inversions. One creates
further linked sources $F \xi$ (where $F$ denotes a product of links) 
with the desired spatial properties, and
computes the quark propagators originating from them, $\phi_F = M^{-1} F
\xi$. Combining the latter with the quark propagator stemming from
$\xi$, i.e.\ $\phi=M^{-1} \xi$, yields the product $\phi^* \phi_F$, from
which, upon averaging over the noise, one can evaluate a set of
correlators with the meson field $\bar{d} \gamma_5 F u$ inserted at the
source (and all possible spatial structures at the sink). Employing
linked sources, as explained above, one can finally evaluate correlators
with the meson field $\bar{d} \Gamma F u$ inserted at the source with
any spatial structure $F$ and Dirac matrix $\Gamma$, while retaining all
advantages of the one-end trick.

In this work we use fuzzing, see Appendix~\ref{subsec:appendixDfuzzing} 
and ref.~\cite{Lacock:1994qx}, to create spatially non-local meson
operators, since this procedure is computationally fast also at the
sink. The fuzzed meson source is constructed from a sum of straight
paths of length $6a$, in the six spatial directions, between quark and
anti-quark. These straight paths are products of fuzzed gauge links.
Here for the fuzzed links we use the iterative procedure defined in
Appendix~\ref{subsec:appendixDfuzzing} with $\lambda_s=0.25$ and $n=5$.

In principle one could hope to extend the approach described above to
baryonic correlators (choosing $\xi$ as a cubic root of $1$) but the
signal to noise ratio will be less favourable (noise induced standard
deviation will be $\approx V_s^{3/2}$ versus signal $\approx V_s$).
Unfortunately one finds that this extension of the stochastic  method to
baryons is not any improvement over using point-like sources. In
general, the choice of the optimal stochastic methods needs to be
investigated on a case by case basis.

\subsection{On the way of choosing the source time-slice}
\label{subsec:source2ways}

As discussed above, we invert  on  spatial-volume stochastic sources
located at time $t_0$, where $0\leq t_0 < T$ can be chosen differently
for each gauge configuration. We have explored two ways of changing the
source time-slice $t_0$. One consists in  moving $t_0$ cyclically
through the lattice. This means that we choose $n$ equally spaced values
for the source time locations, $t_0^{(i)}, 0\leq i<n$.  Then we invert
on the $j$-th gauge configuration using sources that are non-vanishing
only at the time-slices $t_0=t_0^{(j\mod n)}$. Hence, we invert  from the
same time-slice only every $n$ configurations, i.e.\ after one cycle.
Even though this method  should decorrelate the measurement on two
consecutive gauge configurations better than when the time-slices are
kept fixed, it has the drawback that after a relatively short number of
configurations the same time-slice is used again. Actually, at least for
the mesonic correlators studied in this paper, it turns out that two
measurements from the same time-slice, but $8$ trajectories apart, are
much more correlated than two measurements from different time-slices,
but only two trajectories apart. Furthermore the analysis with the
$\Gamma$ method  of ref.~\cite{Wolff:2003sm} described in
sect.~\ref{sec:errors} and Appendix~\ref{sec:appendixC}  becomes ill-defined,
because translational invariance is broken. This invariance can be
recovered, however, by averaging over cycles and using the $\Gamma$
method on the cycle-averaged ensemble.

The second way of moving the time-slice we explored was to choose the
value of $t_0$ randomly for every gauge configuration we inverted
on. This method also maintains translational invariance properly for a
large enough configuration ensemble. It is therefore expected to work 
better than the aforementioned cyclical way.
This will indeed turn out to be the case, as we shall see below, where  
we discuss in more detail the effects of these two ways of generating source
time-slices.

\section{Computations in the neutral meson sector}
\label{sec:neutralmesons}

Lattice QCD with maximally twisted Wilson fermions enjoys the
remarkable property that, even if the action is not $\mathrm{O}(a)$ 
improved, all the physically relevant observables
are affected by cutoff effects only at order $a^2$ (and higher).
Among these $\mathrm{O}(a^2)$ cutoff effects will be a 
violation of parity and (in part) isospin. 
Isospin and parity violations have several consequences for 
meson spectroscopy. For instance 1) neutral and charged mesons can have 
different masses,
2) quark-disconnected contributions are needed for neutral isovector mesons
and 
3) correlators receive contributions from states that in the continuum limit
carry different parity and isospin quantum numbers~\cite{Frezzotti:2003ni}. 
Here we discuss how we compute the correlators for neutral 
mesons and, in particular, the quark-disconnected (for brevity called simply 
``disconnected'' below) contributions.
We illustrate our approach in the relevant case 
of the neutral pseudo-scalar meson. 

The neutral pion can be created by the operator 
$ \sqrt{2} \bar\psi \gamma_5\tau_3 \psi$ which, at maximal twist,
in the twisted quark basis reads
$(i/\sqrt{2}) \bar\chi \chi = (i/\sqrt{2}) (\bar{u} u + \bar{d} d)$.
When this operator is inserted at source
and sink, we will have to consider the correlators
 \begin{equation}
  C_{\rm tot}(t) = 
  \langle (\bar{u}  u +\bar{d} d)(t) ( \bar{u} u +\bar{d} d)(0) \rangle/2\, ,\label{PICORR}
 \end{equation}
where again three-space indices are understood. The latter can be rewritten in the form 
 \begin{eqnarray}
 && C_{\rm tot}(t)=\widetilde{C}(t)+\widetilde{D}(t)\, ,\label{DEF}\\
 &&\widetilde{C}(t)= -\langle {\rm tr}[ G_u(0,t)G_u(t,0)] + 
{\rm tr}[ G_d(0,t)G_d(t,0) ] \rangle/2\, ,\label{TILC}\\
 &&\widetilde{D}(t)= \langle {\rm tr}[  G_u(0,0)+G_d(0,0) ]
                             {\rm tr}[ (G_u(t,t)+G_d(t,t) ]\rangle/2\, ,\label{TILD}
 \end{eqnarray} 
with the trace ${\rm tr}[\dots]$ running only over spin and colour indices.
 As usual, we can relate the connected  contribution ($\widetilde{C}$)
to propagators from a common source (at time $x_0=0$) through
 \begin{equation}
 \widetilde{C}(t)= -\langle  {\rm tr}[ \gamma_5 G_u(0,t)\gamma_5 G_d(0,t)^+] +
                {\rm tr}[ \gamma_5 G_d(0,t)\gamma_5 G_u(0,t)^+] \rangle/2\, .
 \end{equation}
The disconnected contribution can be expressed as 
\begin{equation}
  \widetilde{D}(t)= \langle {\rm tr}[G_u(0,0)+G_u(0,0)^+]
                            {\rm tr}[G_u(t,t)+G_u(t,t)^+] \rangle/2\, .
\end{equation} 
Thus we see that to evaluate the correlation~(\ref{PICORR}) we need
both  $u$ and $d$-quark sources for the connected contribution as well
as  an evaluation of the disconnected contribution for $u$-quarks at
both initial  and final $t$-value. This is at variance with the $\pi^+$
correlator which can be evaluated from  a $u$-quark source alone and
which has no disconnected contribution. The evaluation of the
disconnected loops is detailed in Appendix~\ref{sec:appendixB},
including discussion of both the hopping-parameter method for the
reduction of the stochastic noise~\cite{McNeile:2000xx}  and a new
powerful method of variance reduction applicable  in many cases.

In the table below we give the correspondence between bilinear operators
of the form $\bar{u} \Gamma u \pm \bar{d} \Gamma d$, where $\Gamma$ is
an hermitian combination of $\gamma$-matrices, and the neutral mesonic
state that is associated with each of them in the limit $a \to 0$ (i.e.\
ignoring $\mathrm{O}(a)$ contaminations from states of different parity and
isospin).

\begin{center}
  \begin{tabular}{|c|c|}
    \hline
    Meson & Operator \\
    \hline
    &          \\
    $\pi^0$,\ $\pi^0$,\ $f_0$ &
    $\bar{\chi} i\gamma_0 \gamma_5 \tau_3 \chi$, $\bar{\chi} \chi$, $\bar{\chi} \gamma_5 \tau_3 \chi$ \\
    $\eta$,\ $\eta$,\ $a_0^0$ &
    $\bar{\chi} i\gamma_0 \gamma_5  \chi$, $\bar{\chi} \tau_3 \chi$, $\bar{\chi} \gamma_5  \chi$ \\
    $\rho^0$,\ $\rho^0$,\ $h_1$ &
    $\bar{\chi}\gamma_i\tau_3\chi$, $\bar{\chi}i\gamma_i\gamma_0\gamma_5\chi$, 
                $\bar{\chi}i\gamma_i\gamma_0\tau_3 \chi$ \\
    $\omega$,\ $\omega$,\ $b_1^0$ &
    $\bar{\chi}\gamma_i \chi$, $\bar{\chi} i\gamma_i \gamma_0 \gamma_5 \tau_3 \chi$, 
                $\bar{\chi} i\gamma_i \gamma_0  \chi$ \\
    $a_1^0$ & $\bar{\chi} i\gamma_i \gamma_5 \tau_3 \chi$ \\
    $f_1$ & $\bar{\chi} i\gamma_i \gamma_5  \chi$ \\
    $X_1^0$ & $\bar{\chi}  \gamma_0 \tau_3 \chi$ \\
    $X_0^0$ & $\bar{\chi}  \gamma_0  \chi$ \\
    \hline
    
  \end{tabular}
\end{center}

Here $X_1^0$ ($X_0^0$) labels an isotriplet (isosinglet) state with
$J^{PC}=0^{+-}$, for which no experimental candidate is known. We remark
that these operators are conserved isotriplet  (isosinglet) currents in
the continuum theory with two mass degenerate quarks. 

As in the charged channel, we evaluate the two-point  correlators where
only pairs of meson operators appearing in the same line  of the Table
above are inserted. Since we use a local and extended (fuzzed) source
and sink in each case, we have either $6 \times 6$ or  $2 \times 2$
matrices of correlators available. The connected correlators are
actually the same for certain states of different isospin (e.g.\ $\eta$
or $\pi$). The same `one-end-trick' discussed above, based on the use of
stochastic time-slice  sources with random choice of its position on
each gauge configuration, can be used for the connected neutral
correlator.

In more detail, we use four ``linked'' sources  ($\xi^{(\beta)}$, see
sect.~\ref{subsec:sources}) and further four fuzzed sources based on the
same noise field ($F\xi^{(\beta)}$)  to compute ordinary and fuzzed
$u$-quark propagators from one time-slice to all points. This set of
eight sources is just the same we used to evaluate correlators of
charged mesons. For neutral mesons, we inverted the lattice Dirac matrix
of the $d$-quark  on the same (non-fuzzed) four stochastic sources
($\xi^{(\beta)}$) as above and on the corresponding four stochastic
sources with the lowest possible three-momentum ($2\pi/L$, for
simplicity taken always in the $x$-direction). In principle mesonic
operators with non-zero anisotropic three-momentum have less symmetry
then their counterparts with vanishing three-momentum, implying that
more correlators may take non-zero values. Here we do not evaluate these
additional correlators. We do take care,  however, to distinguish
between the vector meson correlators with three-momentum parallel to
spin and those with three-momentum perpendicular to it. As shown
elsewhere~\cite{McNeile:2002fh}, a study  of the difference between
these correlators can shed some light on the  mixing of $\rho$ mesons
with their decay products ($\pi\pi$).

\section{Simulation algorithm and error analysis}
\label{sec:algo}

In this section we provide details on the algorithms we used to generate
the gauge configurations and information on the methods employed for the estimate
of statistical errors.

\begin{table}[t]
  \centering
  \begin{tabular*}{1.\linewidth}{@{\extracolsep{\fill}}lcccc}
    \hline\hline
    $\Bigl.\Bigr.$Run & $L^3\times T$ & $a\mu_q$ &
    $N_\mathrm{traj}$ & $N_\mathrm{cfg}$ \\
    \hline\hline
    $\Bigl.\Bigr.B_{1a}$ & $24^3\times48$ & 
    $0.0040$ & $5000$ & $2500$ \\
 
    $\Bigl.\Bigr.B_{1b}$ &  & $0.0040$ & $1341$ & $670$ \\

    $\Bigl.\Bigr.B_{1c}$ &  & $0.0040$ & $3380$ & $1690$ \\

    $\Bigl.\Bigr.B_2$ & &  $0.0064$ & $5192$ & $2500$ \\

    $\Bigl.\Bigr.B_{3a}$ & & $0.0085$ & $3753$ & $1876$ \\

    $\Bigl.\Bigr.B_{3b}$ & & $0.0085$ & $940$ & $470$ \\

    $\Bigl.\Bigr.B_4$ & & $0.0100$ & $5000$ & $2500$ \\

    $\Bigl.\Bigr.B_{5a,b}$ & & $0.0150$ & $2500$ & $1250$ \\

    \hline\hline
  \end{tabular*}
  \caption[Summary of all simulation points.]{Summary of all
    simulation points. We give the lattice size 
    $L^3\times T$ and the value
    of the twisted mass $a\mu_q$. In the last two columns we quote the number of
    equilibrated trajectories $N_\mathrm{traj}$ produced and the number
    of configurations $N_\mathrm{cfg}$ saved to disk and finally stored 
    within ILDG, see the review~\cite{Jansen:2006ks} for further links 
    and references. All runs listed in the Table have been performed 
    at $\beta=3.9$ and $\kappa=0.160856$.}
  \label{tab:setup}
\end{table}

In Table~\ref{tab:setup} we give the list of the key parameters characterising the  
simulations we are going to use in this paper.
All simulations $B_{1}-B_{5}$ have been performed at a fixed
value of the gauge coupling $\beta=3.9$ and a fixed value of the hopping
parameter $\kappa=(8+2am_0)^{-1}=0.160856$ on $24^3\times48$ lattices.
In addition to the values of $a\mu_q$
we provide in Table~\ref{tab:setup} the number of trajectories,
$N_\mathrm{traj}$, produced after allowing for $1500$ equilibration
trajectories, and the number of gauge configurations, $N_\mathrm{cfg}$,
that were saved on disk (one every second trajectory). For every value of
$a\mu_q$ we have reached $\sim 5000$ equilibrated trajectories. 

In case we have several ensembles (as for instance for $B_1$) or
several replicas (as for instance for $B_5$) for the same lattice 
parameter set, we denote this by adding 
an extra subscript, $a,b,...$. For our smallest value $a\mu_q=0.004$ we
extended our statistic from about 5000 trajectories (ensemble $B_{1a}$)
to $\sim 10000$ trajectories (if also trajectories from ensembles $B_{1b}$ 
and $B_{1c}$ are counted).

The algorithm we used is a HMC algorithm~\cite{Duane:1987de} with
mass preconditioning~\cite{Hasenbusch:2001ne,Hasenbusch:2002ai} and
multiple time scale integration, as described in detail in
refs.~\cite{Urbach:2005ji,Jansen:2005yp}. The algorithm parameters we
employed for the various runs can be found in Table~\ref{tab:alg-param},
where we mostly follow the notation of ref.~\cite{Urbach:2005ji}.
The integration schemes we used are the Sexton-Weingarten (SW) 
scheme~\cite{Sexton:1992nu}, the second order minimal norm scheme 
(2MN)~\cite{Takaishi:2005tz} and its position version (2MNp). We also list
the number of integration steps $N_i$ for time-scale $i$ (for details see
ref.~\cite{Urbach:2005ji}). We recall that $N_2$ represents the number of
integration steps of the outermost (largest) time-scale. Thus the number of
integration steps of the smallest (i.e.\ innermost) time-scale 
(the one referring to the gauge field integration) is given by $N_2\cdot N_1\cdot N_0$. 
The preconditioning mass is given by $\tilde\mu_1=2\kappa\mu_1$,
with $\mu_1$ typically larger than $\mu_q$ by a factor O(10). 

The second order minimal norm integration scheme on time-scale $i$ is
parametrised by one real number, $\lambda_i$. We also give the
number, $N_i^\mathrm{csg}$, of solutions of the Dirac equation we save
for the chronological solver guess~\cite{Brower:1995vx} 
with the purpose of evaluating the two force terms ($i=1,2$) associated 
to pseudo-fermion integration 
($i=0$ refers to the pure gauge force). 
The notation $N_{1,2}^\mathrm{csg}=0$ means that no chronological solver 
guess was used there. 
Finally, we quote the acceptance rate $P_\mathrm{acc}$ observed
in the simulation and the integrated autocorrelation time
$\tau_\mathrm{int}(P)$ of the plaquette expectation value. The
trajectory length was set to $\tau=1/2$ in all our runs and we always
used $N_\mathrm{PF}=2$ pseudo-fermion fields.
For details on the linear solvers we employed to invert the Dirac
matrix we refer to ref.~\cite{Chiarappa:2006hz}.

To give guidance on the computational cost of such simulations, we 
specify the resource used at our lightest $\mu_q$-value where the CG
iterations for one trajectory cost about $115$ Tflop. The production of
5000 trajectories amounted to about $17$ rack days on the BlueGene/L
installation in J\"ulich, with our code running with an efficiency of
about 18\% for the $B_1$ parameter set. 

\begin{table}[t]
  \centering
  \begin{tabular*}{1.\linewidth}{@{\extracolsep{\fill}}lccccccc}
    \hline\hline
    $\Bigl.\Bigr.$Run & Int. & $N_{0,1,2}$ & $\tilde\mu_1$ &
    $\lambda_{0,1,2}$ & $N^\mathrm{csg}_{1,2}$ & $P_\mathrm{acc}$ &
    $\tau_\mathrm{int}(P)$ \\ 
    \hline\hline
    $\Bigl.\Bigr.B_{1a,b}$ & 2MNp & $2,3,6$ & $0.018$ & $0.19,
    0.20, 0.21$ & $0,0$ & $0.85$ & $47(15)$ \\

    $\Bigl.\Bigr.B_{1c}$ & SW & $2,3,6$ & $0.018$ & $-$ & $0,0$
    & $0.90$ & $43(15)$\\

    $\Bigl.\Bigr.B_{2}$ & 2MNp & $2,3,6$ & $0.025$ & $0.19,
    0.20, 0.21$ & $0,0$ & $0.90$ & $23(7)$ \\

    $\Bigl.\Bigr.B_{3a,b}$ & 2MN & $2,3,5$ & $0.020$ & $0.19,
    0.20, 0.21$ & $7,1$ & $0.90$ & $13(3)$ \\

    $\Bigl.\Bigr.B_{4}$ & 2MNp & $2,3,6$ & $0.035$ & $0.19,
    0.20, 0.21$ & $0,0$ & $0.90$ & $15(4)$ \\

    $\Bigl.\Bigr.B_{5a,b}$ & SW & $2,2,6$ & $0.050$ & $-$ & $0,0$ 
    & $0.90$ & $30(8)$ \\
    \hline\hline
  \end{tabular*}
  \caption[HMC algorithm parameters.]{HMC algorithm parameters. For
    all ensembles we specify the 
    integration scheme, the number of time steps on each time scale
    $N_{0,1,2}$, the precondition mass $\tilde\mu_1=2\kappa\mu_1$, the
    $\lambda$-values for the 2MN integration scheme, the number of saved
    solutions $N_\mathrm{csg}$ for the chronological solver guess, the
    acceptance rate $P_\mathrm{acc}$ observed in the run and the
    integrated autocorrelation time of the plaquette
    $\tau_\mathrm{int}(P)$. The trajectory length was set to $\tau=1/2$
    for all runs and we used always $N_\mathrm{PF}=2$ pseudo-fermion
    fields.  }
  \label{tab:alg-param}
\end{table}

\subsection{Statistical error analysis}
\label{sec:errors}

A reliable estimate of the {\em statistical} errors on the measured quantities
is extremely important for many reasons. We discuss here only the points
which are of special relevance in our analysis. 
If the basic {\em systematic} effects in the lattice simulation, originating 
from the lattice discretization, the finite volume  
and the mass of the dynamical quarks   
are to be addressed, the statistical accuracy on all the relevant 
quantities has to be understood very well. In fact,  
on the one hand, relevant but tiny systematic effects
can only be detected with high statistical accuracy, 
on the other hand underestimated statistical errors 
can artificially increase the significance of 
systematic effects.
The PCAC quark mass, though not a physical quantity, plays here 
a special role, since 
the precision by which it is set to zero is related to
the accuracy (see sects.~\ref{subsec:1_2}--\ref{subsec:1_3}) by which we can expect 
to be at maximal twist. 
Secondly, small statistical errors in low-energy hadronic quantities is an 
expected virtue of the twisted mass formulation, where a sharp infra-red cut-off
ensures a stable MC evolution of the lattice system. Of course 
we have to make sure that an apparently small statistical error does not 
come as a result of large unnoticed autocorrelations 
in the MC history. So autocorrelations in the measured quantities must be
accurately analysed. Finally, a detailed analysis of the statistical errors
delivers as a by-product the integrated autocorrelation time $\tau_{\rm int}$ 
of the studied observable, from which the efficiency of the employed algorithm as a function
of the simulation parameters can be quantified (see sect.~\ref{subsec:autocorr}).

Given the importance of getting a reliable estimate of statistical errors,
results have been cross-checked using 
different approaches. 
As for the estimate of autocorrelation times two different kinds of analyses 
have been performed:
one based on a standard data-blocking (or binning) procedure and 
another one relying on the so-called $\Gamma$-method~\cite{Frezzotti:2000rp,Wolff:2003sm}.
In order to keep self-contained this paper we discuss these methods in some detail
in Appendix~\ref{sec:appendixC}. 
Since there are arguments~\cite{Wolff:2003sm} supporting the 
superiority of the $\Gamma$-method over data-blocking, the former will be our method 
of choice in the evaluation of $\tau_{\rm int}$ and the error on it, for several observables. 
In particular the $\Gamma$-method has been used to estimate the statistical 
error on the plaquette and $m_{\rm PCAC}$, which
turn out to have large autocorrelation times. 
For all the other observables having
significantly smaller autocorrelation times 
data-blocking and $\Gamma$-method typically give quite similar error estimates.

Cross-correlations among different observables are properly taken into account
in our error analysis by using standard jackknife or bootstrap~\cite{EfronTibshirani:1993} 
or performing fits based on a definition of $\chi^2$ that involves the inverse 
covariance matrix (see eq.~(\ref{eq:chi2cov}) and the discussion of Method A 
in sect.~\ref{subsec:staterrxpt}).

\subsection{Autocorrelation times}
\label{subsec:autocorr}

For a primary observable $O$, i.e.\ one that can be viewed as a
linear combination of expectation values of multi-local operators, 
the integrated autocorrelation time is in principle given by
\begin{equation}
  \label{eq:tauint}
  \tau_\mathrm{int}(O)\ =\ \frac{1}{2} +
  \sum_{n=1}^{\infty}\frac{\Gamma_O(n)}{\Gamma_O(0)}\, ,
\end{equation}
where $\Gamma_O(n)$ is the autocorrelation function
of the observable $O$ (see eq.~(\ref{eq:est_gamma})). The
autocorrelation times for the plaquette and fermionic
quantities, like $am_\mathrm{PS}$, $af_\mathrm{PS}$ and
$am_\mathrm{PCAC}$, were determined using the $\Gamma$-method as
described in ref.~\cite{Wolff:2003sm} (see also
Appendix~\ref{sec:appendixC}). This method allows the
determination of $\tau_\mathrm{int}$ also for non-primary quantities,
as the aforementioned fermionic observables. The values for the
plaquette integrated autocorrelation time are collected in
Table~\ref{tab:alg-param}, those for $am_\mathrm{PS}$, $af_\mathrm{PS}$
and $am_\mathrm{PCAC}$ in Table~\ref{tab:autocorr}.
All quoted values are given in units of trajectories of length 1/2. 

In the case of the ensemble $B_{1a}$ we employed
the two ways of moving the stochastic source through the lattice
described in sect.~\ref{subsec:source2ways}. As can be seen
in Table~\ref{tab:autocorr}, indeed the random way performs
better. This is especially significant for $am_\mathrm{PCAC}$,
for which we observe the longest autocorrelation time among the fermionic quantities. 
For $am_\mathrm{PS}$ and $af_\mathrm{PS}$ the difference
between the two methods is not significant. 
The somewhat larger autocorrelation time of $B_2^{cyc}$,
in particular for $m_{\rm PS}$ and $f_{\rm PS}$,
stems presumably 
from the fact that the time slice sources were chosen closer to 
each other than at the other ensembles. 

Table~\ref{tab:method} gives details on the computational methods
employed to extract the various fermionic quantities. 
In the case where the random way of moving the source is used, 
the value of $t_p$ reported there represents the number of trajectories 
(of length 1/2) between two consecutively measured gauge configurations.
We also give in this Table the value of $\chi^2$/d.o.f.\ obtained when 
the charged pion correlators (for Euclidean time separations 
in the range $10 \leq t/a \leq 23$) are fitted using the Ansatz~(\ref{eq:factorisingfit}).

Looking at the three fermionic observables reported in Table~\ref{tab:autocorr}, 
we observe that the integrated autocorrelation times of $am_\mathrm{PCAC}$ are 
significantly larger than those of $am_\mathrm{PS}$ and $af_\mathrm{PS}$. 
We attribute the large value of the autocorrelation time of $am_\mathrm{PCAC}$
to the peculiar phase structure of twisted mass lattice QCD 
with Wilson type quarks as discussed in ref.~\cite{Farchioni:2004us}.
The simulated values of $\mu_q$ are not in a region where the phase
transition occurs. However, the system may  still feel the
presence of this phase transition. The situation is similar for the
plaquette value, as also discussed in ref.~\cite{Farchioni:2004us}, and
indeed the integrated autocorrelation times for the plaquette and
$am_\mathrm{PCAC}$ are rather similar.

\begin{figure}[t]
  \centering
  {\includegraphics[width=0.75\linewidth,angle=-90]{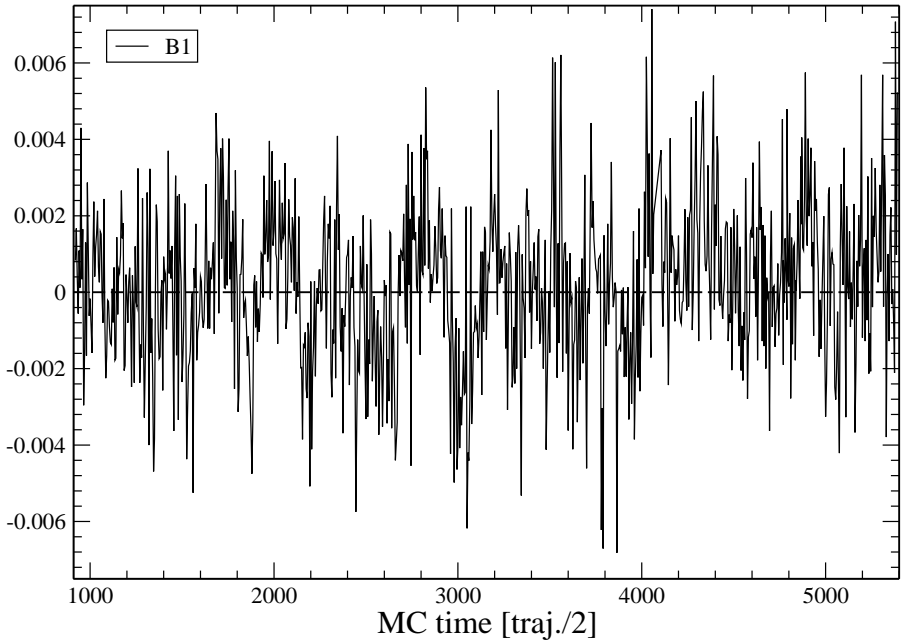}}\quad
 \caption{The Monte Carlo history of the ratio of correlators
defining the PCAC quark mass estimator described in the text
on the configurations of the ensemble $B_1$.
The configuration number corresponds to the
number of trajectories divided by two.}
 \label{fig_algo_mpcac}
 \end{figure}

\begin{figure}[t]
  \centering
  {\includegraphics[width=0.75\linewidth,angle=-90]{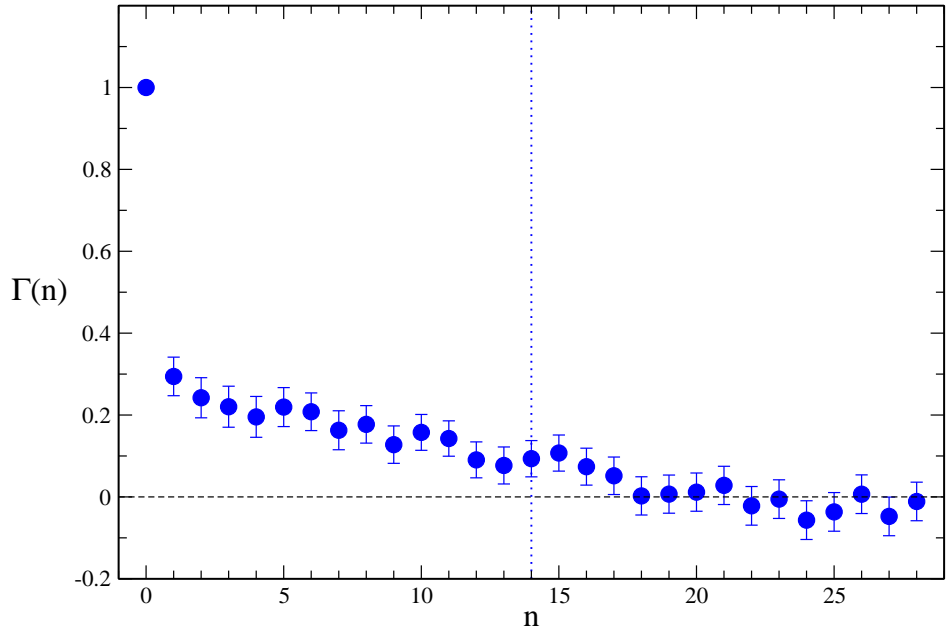}}\quad
 \caption{The autocorrelation function (see~eq.~(\ref{eq:est_gamma})) from the data presented
in fig.~\ref{fig_algo_mpcac}. The vertical line shows
the window $W$ from eq.~(\ref{eq:est_tau_int}) used to evaluate
the integrated autocorrelation function.
 }
 \label{fig_algo_autocorr_mpcac}
 \end{figure}

We show the Monte Carlo (MC) history of 
an estimator of $am_\mathrm{PCAC}$, see eq.~(\ref{eq:mpcac2}), for our
lightest quark mass ($a\mu_q=0.0040$) 
in fig.~\ref{fig_algo_mpcac}. More precisely, we plot for each gauge configuration 
the axial-pseudoscalar correlator at
$t/a=10 $ (where the pion ground state is dominant) multiplied by the factor 
$0.5am_{\mathrm{PS}}/C_{\mathrm{PP}}(10)$, where the average over all 
gauge configurations is used in $C_{\mathrm{PP}}(10)$~\footnote{Note that the average
of this quantity over all gauges is not our best estimator of
$am_\mathrm{PCAC}$, since it does not exploit the possibility
of averaging over $t$ (the Euclidean time separation).}.
The plot in fig.~\ref{fig_algo_mpcac} shows long-ranged
fluctuations in MC time.
The autocorrelation function in~eq.~(\ref{eq:est_gamma}) from this data set is reported in
fig.~\ref{fig_algo_autocorr_mpcac}. 

From it an integrated autocorrelation equal to 32(9) trajectories is obtained, 
in agreement with the 
result quoted in Table.~\ref{tab:autocorr} (third line).

Concerning the neutral pseudoscalar meson, 
a study of the corresponding
correlators indicates that 
the autocorrelations are definitely shorter than 100 
(length 1/2) trajectories. 
Thus our error estimates, coming from a bootstrap analysis on blocked data 
with blocks made of measurements taken from 80 
trajectories, are expected to be reliable. 

Within the relatively large errors of our estimates of the autocorrelation times,
it is actually not possible to find a significant dependence 
on the value of the twisted mass $a\mu_q$ for any of the fermionic
quantities discussed here.

\begin{table}[t]
  \centering
  \begin{tabular*}{1.\linewidth}{@{\extracolsep{\fill}}lccc}
    \hline\hline
    $\Bigl.\Bigr.$Run &
    $\tau_\mathrm{int}(am_\mathrm{PS})$ &
    $\tau_\mathrm{int}(af_\mathrm{PS})$ &
    $\tau_\mathrm{int}(am_\mathrm{PCAC})$ \\
    \hline\hline
    $\Bigl.\Bigr.B_{1a}^\mathrm{cyc}$ & $7(1)$ & $13(4)$ &
    $60(24)$ \\

    $\Bigl.\Bigr.B_{1a}^\mathrm{rnd}$ & $6.6(1.1)$ & $8(1)$ &
    $20(5)$ \\

    $\Bigl.\Bigr.B_{1a,b,c}^\mathrm{rnd}$ & $5.9(7)$ & $7(1)$ &
    $23(5)$ \\

    $\Bigl.\Bigr.B_{2}^\mathrm{cyc}$ & $17(4)$ & $33(8)$ & $43(14)$ \\

    $\Bigl.\Bigr.B_{3}^\mathrm{cyc}$ & $10(2)$ & $11(2)$ & $66(27)$ \\

    $\Bigl.\Bigr.B_{4}^\mathrm{cyc}$ & $7(2)$ & $14(4)$ & $54(23)$ \\

    $\Bigl.\Bigr.B_{5a,b}^\mathrm{cyc}$ & $20(6)$ & $14(3)$ & $105(51)$\\
    \hline\hline
  \end{tabular*}
  \caption[Autocorrelation times of fermionic quantities.]{Estimated
    integrated autocorrelation times for
    $am_\mathrm{PS}$, $af_\mathrm{PS}$ and
    $am_\mathrm{PCAC}$. The labels $cyc$ and $rnd$ refer
    to the cyclic and random choice of the source,
    see text. All integrated autocorrelation times are given
    in units of trajectories of length $1/2$.
    The fact that for the ensemble $B_{5}^\mathrm{cyc}$ we find a rather 
    large autocorrelation time with, however, a large error
    we attribute to the usage of 2 replica in the analysis.
    }
  \label{tab:autocorr}
\end{table}

\begin{table}[t]
  \centering
  \begin{tabular*}{1.\linewidth}{@{\extracolsep{\fill}}lcccc}
    \hline\hline
    $\Bigl.\Bigr.$Run & method &
    $t_p$ & timeslices &  $\chi^2/\mathrm{d.o.f.}$ \\
    \hline\hline
    $\Bigl.\Bigr.B_{1a}$ & cyclic  & $8$ &
    $0,12,24,36$ & $0.12/39$\\

    $\Bigl.\Bigr.B_{1a,b,c}$ & random & $10$ & - & $2.50/39$\\

    $\Bigl.\Bigr.B_{2}$  & cyclic & $16$ & $0,6,12,18,24,30,36,42$ &
    $1.15/39$ \\

    $\Bigl.\Bigr.B_{3}$  & cyclic & $8$ & $0,12,24,36$ &
    $2.38/39$ \\

    $\Bigl.\Bigr.B_{4}$  & cyclic & $8$ & $0,12,24,36$ &
    $1.85/39$ \\

    $\Bigl.\Bigr.B_{5a,b}$  & cyclic & $8$ & $0,12,24,36$ &
    $0.99/39$ \\
    \hline\hline
  \end{tabular*}
  \caption[Measurement methods for fermionic quantities.]
  {Measurement methods for fermionic quantities. The source timeslices
    are either chosen in a cyclic way or randomly. For the
    cyclic way, $t_p$ denotes the number of trajectories between two configurations
    for which the same time-slice was used. For the random way, it
    specifies the number of trajectories between two measured configurations. 
    In the cyclic case we also specify the time-slices where the source was located. Finally,
    the fit range we chose to determine $af_\mathrm{PS}$ and $am_\mathrm{PS}$ was always 10-23
    and the corresponding values of $\chi^2/\mathrm{d.o.f.}$ for a $2 \times 2$ factorising fit 
    (see eq.~(\ref{eq:factorisingfit})) are quoted in the last column.}
  \label{tab:method}
\end{table}

\section{The scale from the static potential}
\label{sec:scale_static_potential}

A convenient way to set the scale in lattice simulations is through measurements 
of the static potential and the associated hadronic scale $r_0$~\cite{Sommer:1993ce}. 
Although we will finally not use $r_0$ to set the scale in our dynamical
simulations, we will use it for a scaling analysis towards the continuum 
limit and its reliable determination is therefore important for us.
The scale $r_0$ is defined via the force between static
quarks at intermediate distance 
\begin{equation}\label{eq:defhadronicscale}
  r_0^2 F(r_0) = 1.65 \, ,
\end{equation}
where numerical calculations are most reliable and hence are expected to lead to very
accurate results. We measure the static quark-antiquark potential by
determining expectation values of Wilson loops of size $r\times t$ on our
ensembles of configurations. Unfortunately, the relative errors of the Wilson
loop expectation values increase exponentially with the temporal extension $t$. 
To reduce these statistical fluctuations one can employ improved static
actions amounting to use modified temporal links for building Wilson loops~\footnote{See 
ref.~\cite{Hasenfratz:2001tw} for a first use of this idea.}. However, it
is also important to enhance the overlap with the physical ground state of the
static system and this can be achieved by invoking iterative spatial smearing
techniques together with a variational method to extract the ground state. The
computational details for calculating the static potential are given in
Appendix~\ref{sec:appendixD} while in the following we want to concentrate 
on analysis details and physical results.

\subsection{Analysis details and results}\label{subsec:analysis_details_scale}

In order to extract the physical scale through eq.~(\ref{eq:defhadronicscale}) 
we need an interpolation of the potential and correspondingly of the
force between the quarks for arbitrary distances $r$. This interpolation is achieved 
by fitting the form of $V(r)$ with the ansatz~\footnote{Note
that we do not use tree level improved distances.} 
\begin{equation}\label{eq:potential_ansatz}
V(r) = V_0 +\frac{\alpha}{r} + \sigma r \, .
\end{equation}
  
We employ a two step procedure to perform the interpolation. First we extract the 
values of the potential $V(r)$ for each $r$ separately using standard variational 
techniques. In a second step we fit directly the potential ansatz in
eq.~(\ref{eq:potential_ansatz}) to the Wilson loop correlators taking
into account all spatial and temporal cross-correlations in the data.
These two steps are now described in more detail (see also~\cite{Niedermayer:2000yx}).

\begin{figure}[t!]
    \begin{center}
      \includegraphics[draft=false,width=0.85\textwidth]{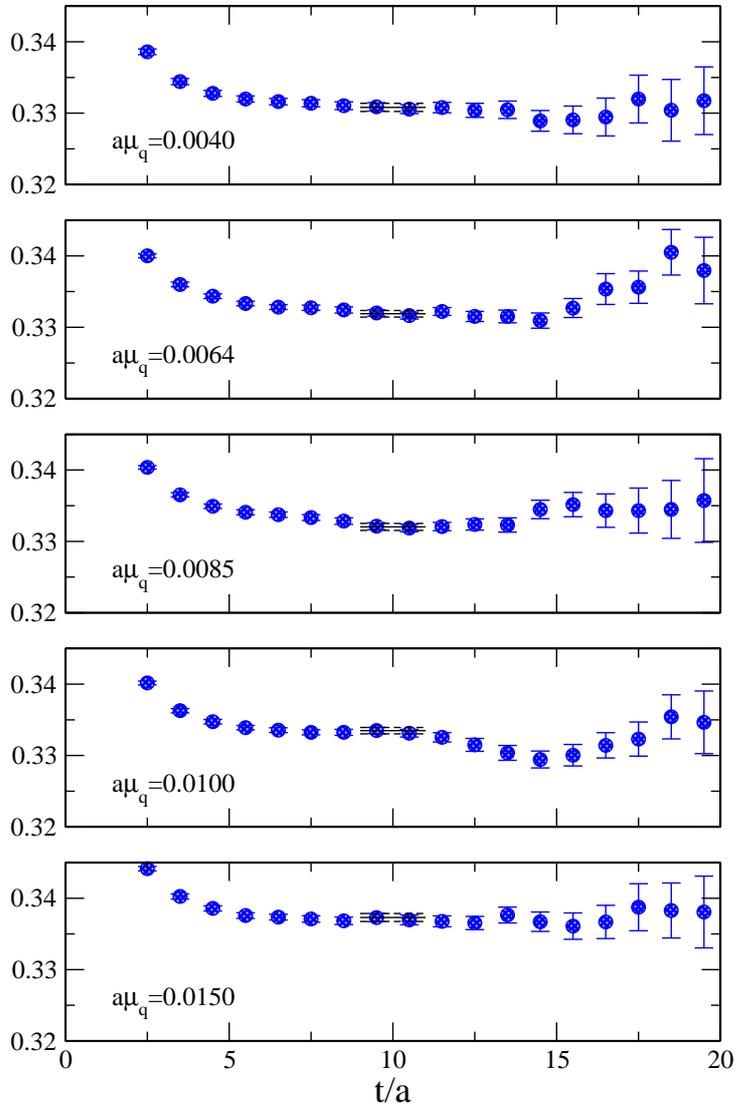}
    \end{center}
\caption{Effective masses for the ground state energy of the Wilson loop at 
  quark-antiquark separation $r/a = 4$, for ensembles $B_1$ to $B_5$.}
\label{fig:eff_mass_r04}
\end{figure}

We use five spatial smearing levels ${\cal S}_n U, n =8,16,24,32,40,$ and hence 
we end up measuring a $5 \times 5$ correlation matrix $C(r,t)$ of spatially smeared
and temporally improved Wilson loops (see Appendix~\ref{sec:appendixD}). The
variational method results in a linear combination of the string operators, 
which projects sufficiently well to the ground
state of the string, i.e.\ has the effect of eliminating the closest excited
string states. 
This is done by solving the generalised eigenvalue problem
\begin{equation}
  C(r,t_1) v_i  = \lambda_i(r;t_0,t_1) C(r,t_0) v_i, \quad \lambda_1
\geq \ldots  \geq \lambda_5 \, ,
\label{eq:GEVP}
\end{equation}
with $t_0=3a$ and $t_1=4a$ and projecting the correlation matrix to the eigenspace
corresponding to the largest eigenvalue, i.e. the ground state,
\begin{equation}
  \bar C(r,t)  = (v_1,C(r,t) v_1) \, .
\end{equation}
Based on effective masses and on a $\chi^2$-test which takes the temporal 
cross-correlations between 
$\bar C(r,t)$ and $\bar C(r,t')$ into account, we choose a plateau region from
$t_{\text{min}}$ to $t_{\text{max}}$. Too small $t$ values distort the
results due to contamination of excited states, while too large values
introduce noise. Examples of effective mass plateaus and the chosen fit ranges
are provided in figure~\ref{fig:eff_mass_r04} for the Wilson loop correlators
of the five ensembles $B_{1,\ldots,5}$ at quark-antiquark separation $r/a = 4$.

The results of our fits are collected in Tables~\ref{tab:Wilson_loop_fit_B1}--\ref{tab:Wilson_loop_fit_B5} 
where we list the plateau regions (fit range), the values of the extracted potential, $V(r)$, and the
$\chi^2$ per degree of freedom. The uncertainties in the extracted values of 
$V(r)$ are calculated using a non-parametric bootstrap method~\cite{EfronTibshirani:1993}.

\begin{table}[h]
  \begin{tabular*}{\textwidth}[l]{c@{\extracolsep{\fill}}cccc}
    \hline\hline
    $r/a$ & $t_\text{min}$ & $t_\text{max}$  & $a V(r)$ & $\chi^2/$d.o.f. \\
    \hline\hline
    4  &  9 & 11  & 0.3308(10) & 1.09 \\
    5  &  8 & 10  & 0.3974(15) & 0.09 \\
    6  &  7 &  9  & 0.4577(26) & 2.48 \\
    7  &  7 &  9  & 0.5126(31) & 1.60 \\
    \hline\hline
  \end{tabular*}
  \caption{Fit parameters for the Wilson loop correlators for run $B_1$.}
  \label{tab:Wilson_loop_fit_B1}
\end{table}

\begin{table}[h]
  \begin{tabular*}{\textwidth}[l]{c@{\extracolsep{\fill}}cccc}
    \hline\hline
    $r/a$  & $t_\text{min}$ & $t_\text{max}$  & $a V(r)$ & $\chi^2/$d.o.f. \\
    \hline\hline
    4  &  9 & 11  & 0.3319(07) & 1.56 \\
    5  &  8 & 10  & 0.3994(09) & 3.27 \\
    6  &  7 &  9  & 0.4589(12) & 0.16 \\
    7  &  7 &  9  & 0.5129(21) & 0.13 \\
    \hline\hline
  \end{tabular*}
  \caption{Fit parameters for the Wilson loop correlators for run $B_2$.}
  \label{tab:Wilson_loop_fit_B2}
\end{table}

\begin{table}[h]
  \begin{tabular*}{\textwidth}[l]{c@{\extracolsep{\fill}}cccc}
    \hline\hline
    $r/a$ & $t_\text{min}$ & $t_\text{max}$ & $a V(r)$ & $\chi^2/$d.o.f. \\
    \hline\hline
    4  & 9  & 11  & 0.3321(08) & 0.75 \\
    5  & 8  & 10  & 0.4002(11) & 4.39 \\
    6  & 7  &  9  & 0.4617(17) & 0.40 \\
    7  & 7  &  9  & 0.5177(22) & 0.02 \\
    \hline\hline
  \end{tabular*}
  \caption{Fit parameters for the Wilson loop correlators for run $B_3$.}
  \label{tab:Wilson_loop_fit_B3}
\end{table}

\begin{table}[h]
  \begin{tabular*}{\textwidth}[l]{c@{\extracolsep{\fill}}cccc}
    \hline\hline
    $r/a$ & $t_\text{min}$ & $t_\text{max}$ & $a V(r)$ & $\chi^2/$d.o.f. \\
    \hline\hline
    4  &  9 & 11  & 0.3335(06) & 1.40 \\
    5  &  8 & 10  & 0.4013(12) & 0.41 \\
    6  &  7 &  9  & 0.4616(15) & 0.37 \\
    7  &  7 &  9  & 0.5177(25) & 0.98 \\
    \hline\hline
  \end{tabular*}
  \caption{Fit parameters for the Wilson loop correlators for run $B_4$.}
  \label{tab:Wilson_loop_fit_B4}
\end{table}

\begin{table}[h]
  \begin{tabular*}{\textwidth}[l]{c@{\extracolsep{\fill}}cccc}
    \hline\hline
    $r/a$  & $t_\text{min}$ & $t_\text{max}$ & $a V(r)$ & $\chi^2/$d.o.f. \\
    \hline\hline
    4  &  9 &  11 & 0.3373(07) & 1.01 \\
    5  &  8 &  10 & 0.4062(08) & 3.21 \\
    6  &  7 &   9 & 0.4692(14) & 1.24 \\
    7  &  7 &   9 & 0.5272(22) & 2.28 \\
    \hline\hline
  \end{tabular*}
  \caption{Fit parameters for the Wilson loop correlators for run $B_5$.}
  \label{tab:Wilson_loop_fit_B5}
\end{table}

This first step allows to determine for each $r$ the value of the potential $V(r)$.
A straightforward strategy to evaluate the scale $r_0/a$ is then to fit the
numbers so obtained with the ansatz~(\ref{eq:potential_ansatz}).
However, one can diminish the errors on the fit parameters by exploiting the fact 
that data at different values of $r$ are correlated. Therefore, in a second step we 
use the ground state projected correlator $\bar C(r,t)$ to estimate the covariance matrix 
$$ \text{Cov}(r,t;r',t') \; \equiv \; \langle \bar C(r,t) \bar C(r',t')  \rangle 
 \; - \; \langle \bar C(r,t) \rangle \langle \bar C(r',t')  \rangle $$ 
from the bootstrap samples of $\bar C(r,t)$ and use $\text{Cov}(r,t;r',t')$ to construct 
the $\chi^2$ function (see the discussion in sect.~\ref{subsec:staterrxpt}). 
The $r$ and $t$ dependence of $\bar C(r,t)$ is fitted with the formula 
(see eq.~(\ref{eq:potential_ansatz}))
\begin{equation}\label{eq:correlation_fit_ansatz}
\bar C(r,t)\sim Z(r) \exp\left[-tV(r)\right]=
Z(r) \exp\left[-t\left(V_0 + \alpha/r + \sigma r\right)\right]\, .
\end{equation}
For the temporal fit interval we use the fit ranges
$t_{\text{min}}(r)$ to $t_{\text{max}}(r)$ determined in the first step. The
fit range in $r$ is chosen so as to include only a few values of $r$ closest to
$r_0$ in order to minimise both the statistical error and the
systematic error coming from the choice of the interpolation formula.
Once the best fit parameters ($V_0$, $\alpha$ and $\sigma$) in
eq.~(\ref{eq:correlation_fit_ansatz}) are found, the value of $r_0/a$
is obtained straightforwardly by computing the static force from the
derivative w.r.t.\ $r$ of eq.~(\ref{eq:potential_ansatz}) and imposing the
condition~(\ref{eq:defhadronicscale}) that defines $r_0$.

A compilation of the results of our fits is provided in Table~\ref{tab:r0_fit_results} where
we give the number of measurements $N_\text{meas}$ and
the $\chi^2$ per degree of freedom, in addition to the results for $r_0/a$. The
final error on $r_0/a$ is estimated through jackknife and bootstrap procedures
using binning to take residual autocorrelations into account. 
In Table~\ref{tab:r0_fit_results} we give the errors from the jackknife procedure using
a binning factor equal to 4.

\begin{table}[h]
  \begin{tabular*}{\textwidth}[l]{c@{\extracolsep{\fill}}cccc}
    \hline\hline
    Run & $N_\text{meas}$ & d.o.f. & $\chi^2/$d.o.f. & $r_0/a$ \\
    \hline\hline
    $B_1$ & 625 &  5 & 1.44 & 5.196(28)  \\
    $B_2$ & 695 &  5 & 1.80 & 5.216(27)  \\
    $B_3$ & 598 &  5 & 2.58 & 5.130(28)  \\
    $B_4$ & 602 &  5 & 0.57 & 5.143(25)  \\
    $B_5$ & 645 &  5 & 1.92 & 5.038(24)  \\
    \hline\hline
  \end{tabular*}
  \caption{Results of the fits for the scale $r_0/a$  from the static
    potential. The fit range was always $r/a = 4 - 7$. The number 
    of measurements, $N_\text{meas}$, and $\chi^2$/d.o.f.
    are also reported.} 
  \label{tab:r0_fit_results}
\end{table}

\subsection{Discussion}

There are several sources of systematic effects which can distort a precise
and accurate determination of the scale $r_0/a$. Here we would like to discuss
a few checks that we have performed in order to asses these systematic effects
and some procedures to minimise their influence.
\\

\begin{figure}[t]
    \begin{center}
      \includegraphics[draft=false,angle=-90,width=0.9\textwidth]{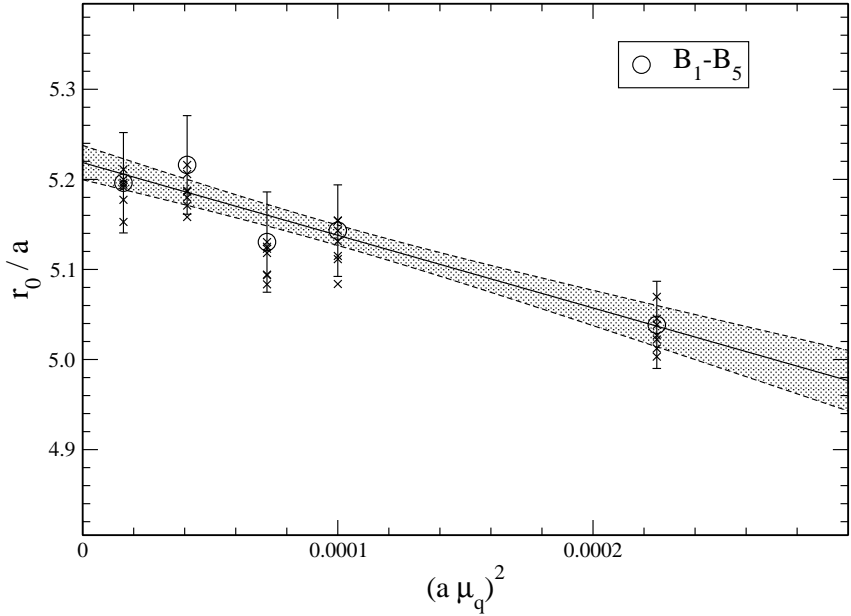}
    \end{center}
\caption{Mass dependence of $r_0/a$. The shaded area shows the error band of
  the quadratic fit (full line) to the data (circles). The additional plus symbols
  are further determinations of $r_0/a$ corresponding to different values of the fit 
  parameters to the ansatz~(\ref{eq:correlation_fit_ansatz}). The spread provides
  an indication of the systematic error due to interpolation (see text) in $r_0/a$.}
\label{fig:r0_results}
\end{figure}

\noindent
{\it Excited states}\\
First of all there are contaminations of the ground state energy of the Wilson
loops from excited states. We expect that these should be eliminated by our
variational calculation of the ground state and our choice of the fit range in
$t$, and we have carefully checked the stability of the results under
variation of the fit parameters (see figure~\ref{fig:r0_results}). In
particular we have checked that we can resolve the first excited state
and that the ground state energy remains stable under this procedure. Moreover
we have also checked the stability of the ground state under a truncation of
the variational operator basis. We would also like to point out that the fit
ranges in $t$ were not chosen independently for each value of $\mu_q$ and $r$,
rather we chose them after taking a global view of the
effective mass data for all values of $\mu_q$ at given fixed $r$
(see fig.~\ref{fig:eff_mass_r04} for the case $r=4a$). 
This procedure makes sense
since the $\mu_q$-dependence of the Wilson
loop correlators is expected to be rather weak (see below) and 
is particularly useful in cases where the choice of the fit range for the
effective masses cannot be determined unambiguously given the available statistics. 
Finally we note that contaminations from excited states tend to increase the 
potential energies and the effect will be more pronounced for the larger Wilson 
loops. As a consequence, residual contributions from excited states will tend to 
decrease the value of $r_0/a$.
\\

\noindent
{\it Interpolation error}\\
The interpolation of the potential (or the force) as a function of $r$ is not
unique. Here we would like to emphasise that we use eq.~(\ref{eq:potential_ansatz}) 
only locally as a simple interpolation ansatz without attaching to it any special 
physical meaning. As a check of this interpolation ansatz, one can 
use separately the matrices of correlators computed for $r/a=4-6$ and for $5-7$ 
to obtain two different
determinations of $r_0/a$. Their difference then provides an estimate of the error
coming from the interpolation procedure. It turns out that our choice of the fit range
$r/a=4-7$ covers this spread typically within 1--2 standard deviations of our final
result (see figure~\ref{fig:r0_results}). 
\\

\noindent
{\it Correlations}\\
We have already pointed out that it is important to take both the spatial and
temporal cross-correlations of the Wilson loop operators into account when fitting
them to the ansatz~(\ref{eq:correlation_fit_ansatz}). Our finite statistics limits
ourselves to short fit ranges in order to obtain a stable covariance
matrix, and this is one of the motivations for the rather narrow fit ranges in
$t$ in Tables~\ref{tab:Wilson_loop_fit_B1}--\ref{tab:Wilson_loop_fit_B5}.  In
order to assess the effect arising from Wilson loop autocorrelations, we form
bins of the data of various sizes, though this reduces the amount of data
available for estimating the covariance matrix even further. In fact, it turns
out that the fits become unreliable beyond bin size 4 and before the binning
error becomes stable. As a consequence we cannot exclude that the errors on
$r_0/a$ are somewhat underestimated due to residual autocorrelations. 
\\

\noindent
{\it Mass dependence}\\
Our results for $(r_0/a)$ are plotted in figure~\ref{fig:r0_results}. We note
that the $a\mu_q$ dependence appears to be rather weak, and hence we expect
the data for the (purely gluonic) observable $r_0/a(a\mu_q)$ to be well
described by polynomials of low order in $a\mu_q$.  In
Table~\ref{tab:r0_vs_mu_fits} we collect the results obtained by fitting our
data at different values of $a\mu_q$ (see
Table~\ref{tab:r0_fit_results}) to few simple functional forms, namely
\begin{equation*}
\begin{array}{rl}
\textrm{(I)}:   &  r_0/a + c_2 (a \mu_q)^2\, , \\
\textrm{(II)}:  & r_0/a + c_1 (a \mu_q)\, , \\
\textrm{(III)}: & r_0/a + c_1 (a \mu_q) + c_2 (a \mu_q)^2 \, .
\end{array}
\end{equation*}
The ansatz (I) is inspired
by the fact that with maximally twisted (unlike the case of untwisted)
Wilson quarks the lattice fermionic determinant of the $N_f=2$ theory depends
only quadratically on the bare quark mass.  A weaker dependence on the bare
quark mass can only appear via the effects of spontaneous chiral symmetry
breaking on the static quark potential and would actually be a 
dependence~\footnote{We are indebted to R.~Sommer for very useful discussions 
on this point.} 
on $|a\mu_q|$.
This is the motivation for the fit ansatz (II), if it can be
assumed that $a\mu_q$ is sufficiently small to make the
$(a\mu_q)^2$-dependence negligible, and (III), if the
$(a\mu_q)^2$-dependence is instead statistically significant.

The fit based on the ansatz (I) describes our data rather well, as shown in
figure~\ref{fig:r0_results}, suggesting that possible effects of spontaneous chiral
symmetry breaking in the static potential at distances around $0.5$~fm are
negligible within our statistical errors. This interpretation is supported also
by the other two fits: even if a $\mu_q$-dependence of the type (II) 
cannot be ruled out completely, we observe that not only the $\chi^2/{\rm d.o.f.}$
of the fit (I) compared to (II) is better, 
but also the best-fit values of $c_2$ from fits (I) and (III) are more consistent
between themselves (and less consistent with zero) than the best-fit values 
of $c_1$ coming from fits (II) and (III). 
We would like to note that these findings are corroborated by analogous fits
of the $a\mu_q$ dependence of the static potential at fixed values of $r/a$, i.e.\ in
situations where no interpolation in $r/a$ is involved.

\begin{table}[h]
  \begin{tabular*}{\textwidth}[l]{c@{\extracolsep{\fill}}cccccccc}
    \hline\hline
    $r_0/a$   & $c_1 \times 10^{-2}$ & $c_2\times 10^{-4}$ & fit range    & $\chi^2/{\rm d.o.f.}$ \\
    \hline
    5.22(2) &  --             & -0.08(2) & $B_1-B_5$ &  0.85 \\
    5.22(3) &  --             & -0.09(4) & $B_1-B_4$ &  1.26 \\ 
    \hline
    5.28(3) &  -0.16(3)          &  --   & $B_1-B_5$  &  1.10 \\
    5.26(5) &  -0.12(6)          &  --   & $B_1-B_4$  &  1.37 \\
    \hline
    5.22(8) &  -0.01(18)      & -0.08(9) & $B_1-B_5$ &  1.28 \\ 
    \hline\hline
  \end{tabular*}
  \caption{Results of the fits of the $a\mu_q$ dependence of $r_0/a$ according to the ansatz
   (I), (II) and (III) in the text.}
  \label{tab:r0_vs_mu_fits}
\end{table}

We conclude that the mass dependence is well described by the ansatz (I) and
remark that an almost identical central value for $r_0/a$ at the chiral point
is obtained from the ansatz (III), which also allows for a linear term in
$a\mu_q$. The ansatz (II)  gives a central value for $r_0/a$ at
the chiral point lying two standard deviations above that from the ansatz (I).
Finally we note that, if the ansatz (I) for the $\mu_q$-dependence of $r_0/a$ is
used, the relative statistical accuracy of our determination of $r_0/a$ in
the chiral limit is better than 1\%.\\

\section{Some selected results}
\label{sec:results}

In this section we present results for quantities related to the
pseudoscalar (PS) channel. This includes, apart from charged and neutral
PS masses and decay constants, also the renormalization
constant $Z_{\rm V}$, which is specifically relevant to maximally twisted mass QCD.

\subsection{Charged and neutral pseudoscalar masses}
\label{ssec:meson}

\vspace{0.3cm}
\noindent\emph{Charged pseudoscalar meson mass}

To extract the charged PS mass $m_\mathrm{PS}$ we consider
the correlation functions discussed in sect.~\ref{subsec:sources}.
We refer to this section and sect.~\ref{sec:errors}
for a detailed discussion of how the correlation
functions are evaluated and the errors are estimated. The
results for the charged PS masses can be found in
Table~\ref{tab:resmpsfps}.

In order to make the effect of the longest runs at $\mu=0.004$ visible,
we quote the results for run $B_{1a}$ and the complete run $B_1$
separately. While for $B_{1a}$ we have $1811$ measurements made in the
cyclic way explained in sect.~\ref{subsec:autocorr}, there are $895$
measurements for $B_1$ performed moving the source time-slice randomly
through the lattice. Even though in the latter case we have 
fewer measurements, they are more decorrelated because the single
measurements are more separated in Monte Carlo time and because the
distance of the position of the sources 
in Euclidean time for two consecutive measurements
is on average larger. It is reassuring to see that results and errors are
consistent between the two sets of data within errors.
{}From this comparison it is also clear that moving the source time-slice
randomly through the lattice is the most convenient of the two methods.

\begin{figure}[t]
  \centering
  {\includegraphics[width=0.75\linewidth,angle=-90]{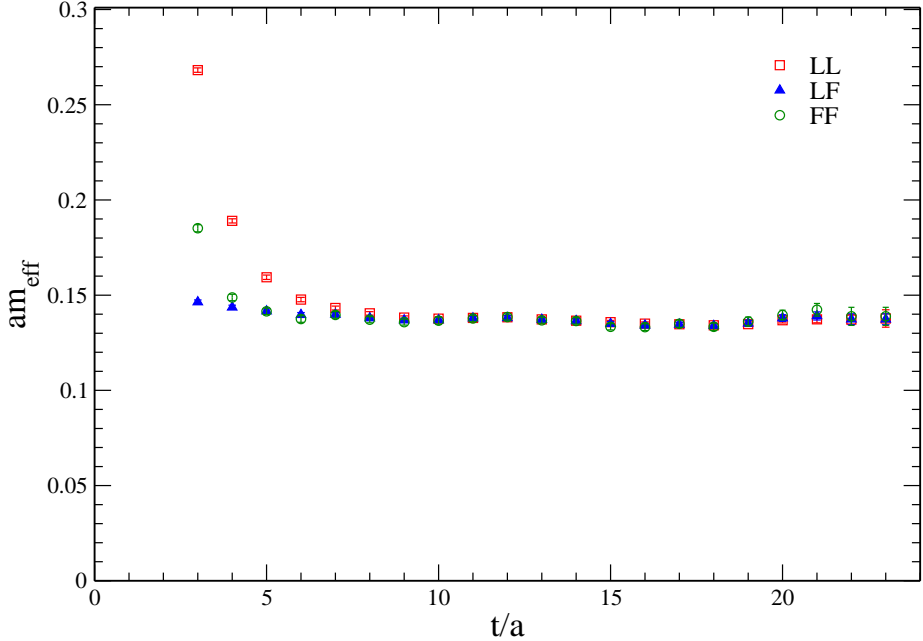}}\quad
  \caption{Effective mass for the pseudoscalar channel from $B_1$ lattice
   data. The effective masses obtained using 3 different interpolating
    operators as described in the text are shown.}
  \label{fig:eff_mass_PS}
\end{figure}

In fig.~\ref{fig:eff_mass_PS} we show examples for effective masses
in the PS channel at our lightest quark mass, extracted from the
PS correlation function (with insertion of the $\bar d\gamma_5 u$
operator) only. We plot the data for the three different choices of the
interpolating operators, namely local-local, local-fuzzed, and
fuzzed-fuzzed. One can see in fig.~\ref{fig:eff_mass_PS} that the
three different operators give compatible results from $t/a\approx10$
on. Hence we are confident that the ground state energy dominates
for $t/a>9$ and we chose the fit range accordingly.

We also attempted to determine the energy of the  first excited state
of the PS meson from a 2-state fit to the 6$\times$6 matrix of
correlators. Even though we were unable to determine the first excited level
in a reliable way from an unconstrained fit, fixing it to the theoretical
value (3 times the ground state mass), as expected in the continuum limit,
does allow an acceptable fit.

This result is quite interesting, as with maximally twisted Wilson
quarks one expects on general grounds also an $\mathrm{O}(a^2)$ contamination 
from the $\pi^0({\bf 0})\pi^\pm({\bf 0})$ two-pion state. 
Such a contamination becomes negligible, 
if compared to the expected three-pion state one in the continuum limit 
(taken at fixed quark mass). It should also be observed that when the 
pion mass is decreased, the two-pion contribution remains negligible 
with respect to the three-pion one until $a^2\Lambda_{\rm QCD}^2 \ll e^{-m_{\rm PS} t}$. 
{}For the range of $m_{\rm PS}$ and $t$ values relevant for our data we find 
that two-pion contamination effects can hardly be detected despite 
our (small) statistical errors (see fig.~\ref{fig:eff_mass_PS}).

\begin{table}[!t]
  \begin{center}
    \begin{tabular*}{1.\linewidth}{@{\extracolsep{\fill}}lcccccc}
      \hline\hline
      Run & $a \mu_{\rm q}$ & $a m_{\rm PS}$ & $a f_{\rm PS}$ &
      $a m_{\textrm{PCAC}}$ & $Z_\mathrm{V}(\mu_q)$ \\
      \hline\hline
      $B_{1a}$ & $0.0040$ & $0.13587(68)$ & $0.06531(40)$
      &$-0.00001(27)$  & $0.6114(85)$ \\
      $B_1$    & $0.0040$ & $0.13623(65)$ & $0.06459(37)$
      &$+0.00017(17)$  & $0.6136(19)$ \\
      $B_2$    & $0.0064$ & $0.16937(36)$ & $0.07051(35)$
      &$-0.00009(17)$  & $0.6096(21)$ \\
      $B_3$    & $0.0085$ & $0.19403(50)$ & $0.07420(24)$
      &$-0.00037(20)$  & $0.6115(22)$ \\
      $B_4$    & $0.0100$ & $0.21004(52)$ & $0.07591(40)$
      &$-0.00097(26)$  & $0.6209(25)$ \\
      $B_5$    & $0.0150$ & $0.25864(70)$ & $0.08307(34)$
      &$-0.00145(42)$  & $0.6165(22)$ \\
      \hline
    \end{tabular*}
  \end{center}
  \caption{Results for masses and decay constants in the charged pseudoscalar
    channel, PCAC quark mass and $Z_\mathrm{V}$. The results for
    the first three quantities come from a fit to a $4\times4$
    submatrix with operators $\bar d\gamma_5 u$ and $i\bar
    d\gamma_0\gamma_5 u$, while for $Z_{\rm V}$ we used the full $6\times6$
    matrix. Note that the difference between the first two rows is
    just the length of the simulation. The time range of the fit was
    always $10-23$ and the $\chi^2/\mathrm{d.o.f.}$ was always smaller than
    one.}
  \label{tab:resmpsfps}
\end{table}

\vspace{0.3cm}
\noindent\emph{Neutral pseudoscalar meson mass}

As discussed in sect.~\ref{sec:neutralmesons}, the neutral PS meson
can be created by interpolating fields that at maximal twist and in the
twisted basis are of the form $\bar{\chi} \chi$ and $\bar{\chi}
\gamma_0\gamma_5 \chi$.
 We evaluate the  correlator (both quark-connected and
quark-disconnected pieces) with each of these operators at source and
sink (also with local and fuzzed variants, thus giving a $4 \times 4$
matrix of correlators) as described above and in
Appendix~\ref{sec:appendixB}. We fit this correlator matrix to one or
more states in the usual way.  Based on our study of autocorrelations
(see sect.~(\ref{sec:algo})), we compute statistical errors by a
bootstrap analysis on blocked data where each 
block includes measurements taken on configurations corresponding 
to a segment of MC history 80 trajectories long. 

\begin{figure}[t!]
  \centering
  \includegraphics[width=0.75\linewidth,angle=-90]{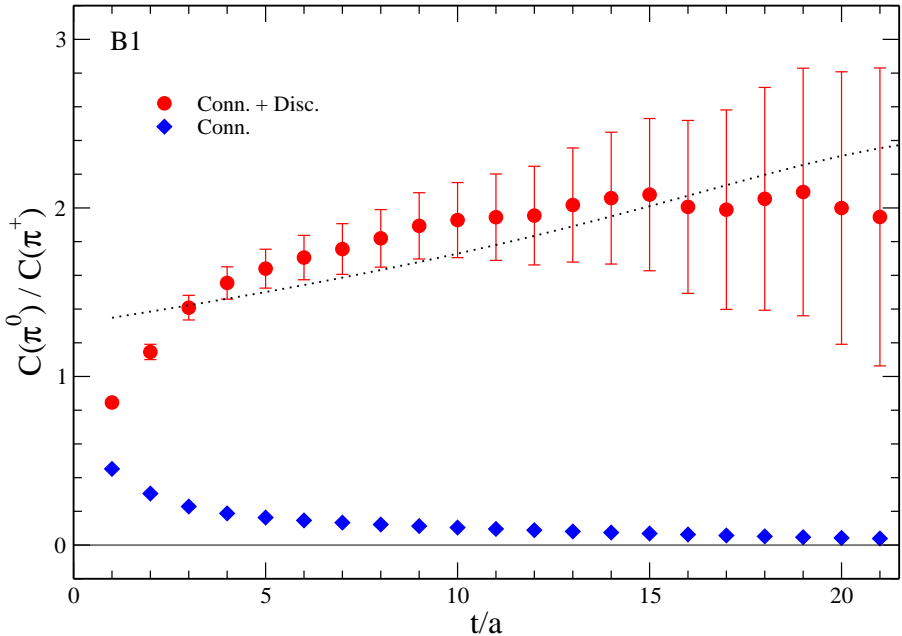}
  \caption{Ratios of correlators of neutral to charged pseudoscalar
    (PS) mesons. The connected  component of the neutral meson correlator is
    shown, as well as the total, which includes the  disconnected
    contribution. The dotted curve illustrates the behaviour of
    the ratio of the correlator as obtained from
    our fitted values of the charged and neutral PS masses.
  }
  \label{fig.disc1}
\end{figure}

Our  results for the neutral PS meson are shown in Table~\ref{tab.disc1}.
Compared to ref.~\cite{Boucaud:2007uk}, we have increased statistics at $\mu_q=0.004$
and we have employed the more refined fitting procedure explained above.
In particular we used $4 \times 4$ matrix of correlators rather than a
$2 \times 2$ matrix. We also include results at a second $\mu_q$ value,
$\mu_q=0.0085$. In order to show the contribution of the
quark-disconnected component to the neutral PS meson mass determination,
we show appropriate ratios in fig.~\ref{fig.disc1}.

\begin{table}
  \centering
  \begin{tabular*}{.9\linewidth}{@{\extracolsep{\fill}}lccccc}
    \hline\hline
    run &$a\mu_q$& $N_\mathrm{meas}$ &  $am^0_\mathrm{PS,conn}$ &
$am^0_\mathrm{PS}$ &
    $af^0_\mathrm{PS}/Z_\mathrm{A}$ \\
    \hline\hline
    $B_1$ & $0.0040$ & $888$ & $0.212(3)$ & $0.109(7)$ & $0.089(3)$\\

    $B_3$ & $0.0085$ & $249$ & $0.259(3)$ & $0.169(11)$ & $0.106(4)$ \\

    \hline\hline
  \end{tabular*}
  \caption{Neutral pseudo scalar meson masses and decay constants at
$\beta=3.9$
    measured from every 10 trajectories at $\mu_q=0.004$
    and every  20 at $\mu_q=0.0085$, as indicated; $am^0_\mathrm{PS,conn}$
is the mass extracted
    from the quark-connected correlators only.
}
  \label{tab.disc1}
\end{table}

We have also evaluated the energies of neutral PS mesons with momentum
$2 \pi/L$ (recall $L/a=24$), obtaining (by use of the continuum dispersion
relation $E^2 = (2\pi/L)^2 + m^2$) mass values consistent with
those shown in Table~\ref{tab.disc1}.

The non-zero momentum results
have the advantage that  no vacuum subtraction is needed for the neutral
PS meson correlator and this provides a crosscheck of the approach we 
employed. For example, at $\mu_q=0.004$ we obtain an energy of
0.309(27) which correspond to a mass 0.164$^{+47}_{-60}$ in lattice units.

\vspace{0.3cm}
\noindent\emph{Pseudoscalar meson mass splitting and related topics} 

Concerning the PS meson mass, it is well known that with maximally
twisted Wilson fermions, even in the theory with $N_f=2$ degenerate
quarks we consider here, there is difference of order $a^2$ (at fixed
quark mass) between the neutral and the charged PS meson mass. Moreover
the latter is very mildly affected by cutoff effects, once maximal
twist is implemented in the optimal way of sect.~\ref{sec:maximaltwist},
as it follows from the formula
$m_{\rm PS}^2 - m_{\rm PS}^2|_{\rm cont}={\mathrm{O}}(a^2\mu_q)+{\mathrm{O}}(a^4)$ 
proved in refs.~\cite{Sharpe:2004ny,Frezzotti:2005gi,Frezzotti:2007qv}. Finally a 
lattice chiral perturbation theory analysis (see 
e.g.\ refs.~\cite{Scorzato:2004da,Sharpe:2004ny,Munster:2004am}) shows that
in the small $\mu_q$ region the difference between the squared neutral and
charged PS masses tends to an $\mathrm{O}(a^2)$ quantity
 \begin{equation}
  r_0^2 ( (m^0_{PS})^2-(m_{PS})^2 ) \simeq c (a/r_0)^2\, ,
  \label{eq.disc1}
\end{equation}
which can be related to one coefficient (usually called $c_2$) of the chiral
effective Lagrangian of (twisted and untwisted) Wilson fermion lattice QCD.
{}From our results we estimate $c= -5.0(1.2)$ and $c = -6.7(2.8)$ respectively
at $\mu_q=0.004$ and $\mu_q = 0.0085$. These determinations are consistent
within errors, as expected. 
Moreover, unlike in the case of lattice formulations with different gluonic
actions~\cite{Farchioni:2005hf}, the sign of $c$ is now in agreement with the 
first order phase transition scenario~\cite{SiSh,Munster:2004am,Sharpe:2004ps},
as predicted by lattice chiral perturbation theory.

Some results for the coupling of the neutral PS meson field to the
divergence of the neutral axial current are also given in
Table~\ref{tab.disc1}. The values of
$f^0_\mathrm{PS}$ are found to be quite close to those obtained for the
charged PS meson (see $f_\mathrm{PS}$ in Table~(\ref{tab:resmpsfps})), if the estimate
$Z_A = 0.76(2)$~\cite{Dimopoulos:2007fn,ETMC:wip} is employed. The
differences between the central values are compatible
with zero within $1$~to~$1.5$ standard deviations of $f^0_\mathrm{PS}$ 
(which is the least precise result of the two).

This good consistency between the two channels is in striking contrast with
the large cutoff effect present in $m_\mathrm{PS}^0$, which we observe as
a large difference (about 50~MeV at $a\mu_q = 0.0040$ and $a\mu_q=0.0085$)
between its value and that of its charged counterpart, $m_\mathrm{PS}$.
Theoretical arguments providing an explanation for the
reasons why a large lattice artifact affects only $m_\mathrm{PS}^0$
were presented in ref.~\cite{Frezzotti:2007qv} and will be further
discussed in a forthcoming publication~\cite{ETMC:Oa2}.

\subsection{PCAC mass}
\label{ssec:pcac}

\begin{figure}[t]
  \centering
  \includegraphics[width=0.75\linewidth,angle=-90]{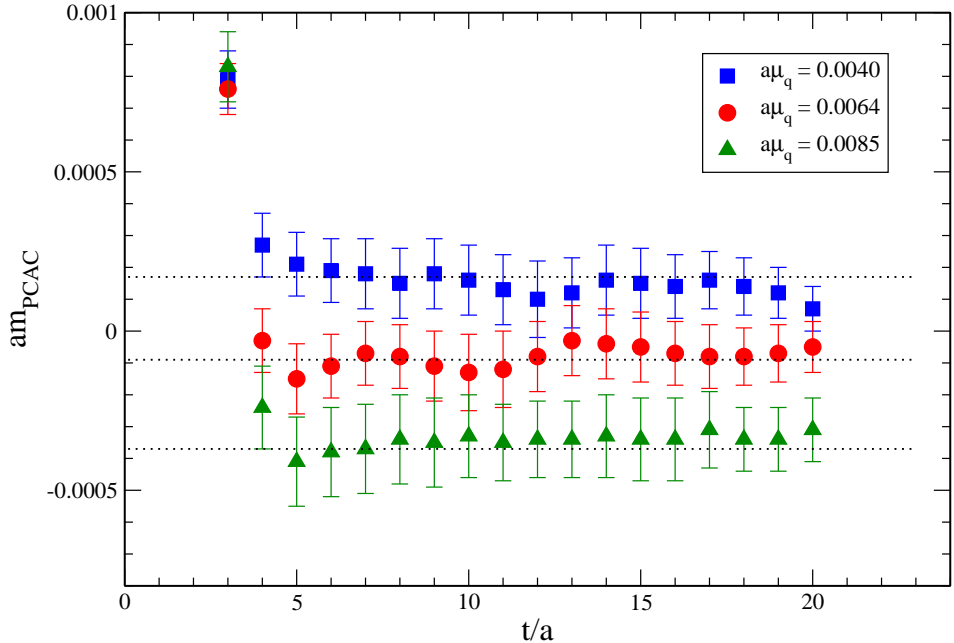}
 \caption{PCAC quark masses versus time separation (from eq.~\ref{eq:mpcac})
for three values of the mass parameter, $\mu_q=0.004$, $\mu_q=0.0064$
and $\mu_q=0.0085$. The lines represent the central values of
the  PCAC quark mass from a fit to a $4 \times 4$ matrix of observables
for a $t$-range 10-23. 
 }
 \label{fig:eff_mass_PCAC}
 \end{figure}

As discussed in sect.~(\ref{sec:maximaltwist}),
we attempt to tune the value of $m_\mathrm{PCAC}$
(see eq.~(\ref{eq:mpcac})) to zero at our minimal $\mu_q$ value,
namely at $a\mu_q=0.004$.
A definition of $m_\mathrm{PCAC}$ equivalent to eq.~(\ref{eq:mpcac})
for time separations so large that the lowest PS meson state dominates, is
given by
\begin{equation}
  \label{eq:mpcac2}
  m_\mathrm{PCAC} = \frac{m_\mathrm{PS}}{2} \frac{\langle0|A_0^a|PS
    \rangle}{\langle0|P^a|PS\rangle}\ , \qquad a=1,2 \, .
\end{equation}
 These two  matrix elements can be directly determined from a fit
to the $4\times4$ matrix involving the interpolating operators $\bar
d\gamma_5 u$ and $i\bar d\gamma_0\gamma_5 u$
(or from the fit to the full $6\times6$
matrix, see sect.~\ref{sec:chargedmesons}). 
The results we obtain when a $4\times4$ matrix is used
are summarised in Table~\ref{tab:resmpsfps} and shown as a horizontal line 
in fig.~\ref{fig:eff_mass_PCAC}.  It is important to notice
that the condition discussed around eq.~(\ref{eq:relprec}) is fulfilled
for all our simulation points, and in particular for
$a\mu_q=0.004$. 

In fig.~\ref{fig:eff_mass_PCAC}, we also illustrate the time dependence
of the local determination of the PCAC quark mass through
eq.~(\ref{eq:mpcac}).
We see 
that the values of $m_\mathrm{PCAC}$ determined using
eq.~(\ref{eq:mpcac}) in this way and eq.~(\ref{eq:mpcac2}) agree very
well between themselves, in the $t$-region where the ground state 
pseudoscalar meson dominates, as expected.

\begin{figure}[t]
  \centering
  \includegraphics[width=.75\linewidth,angle=-90]{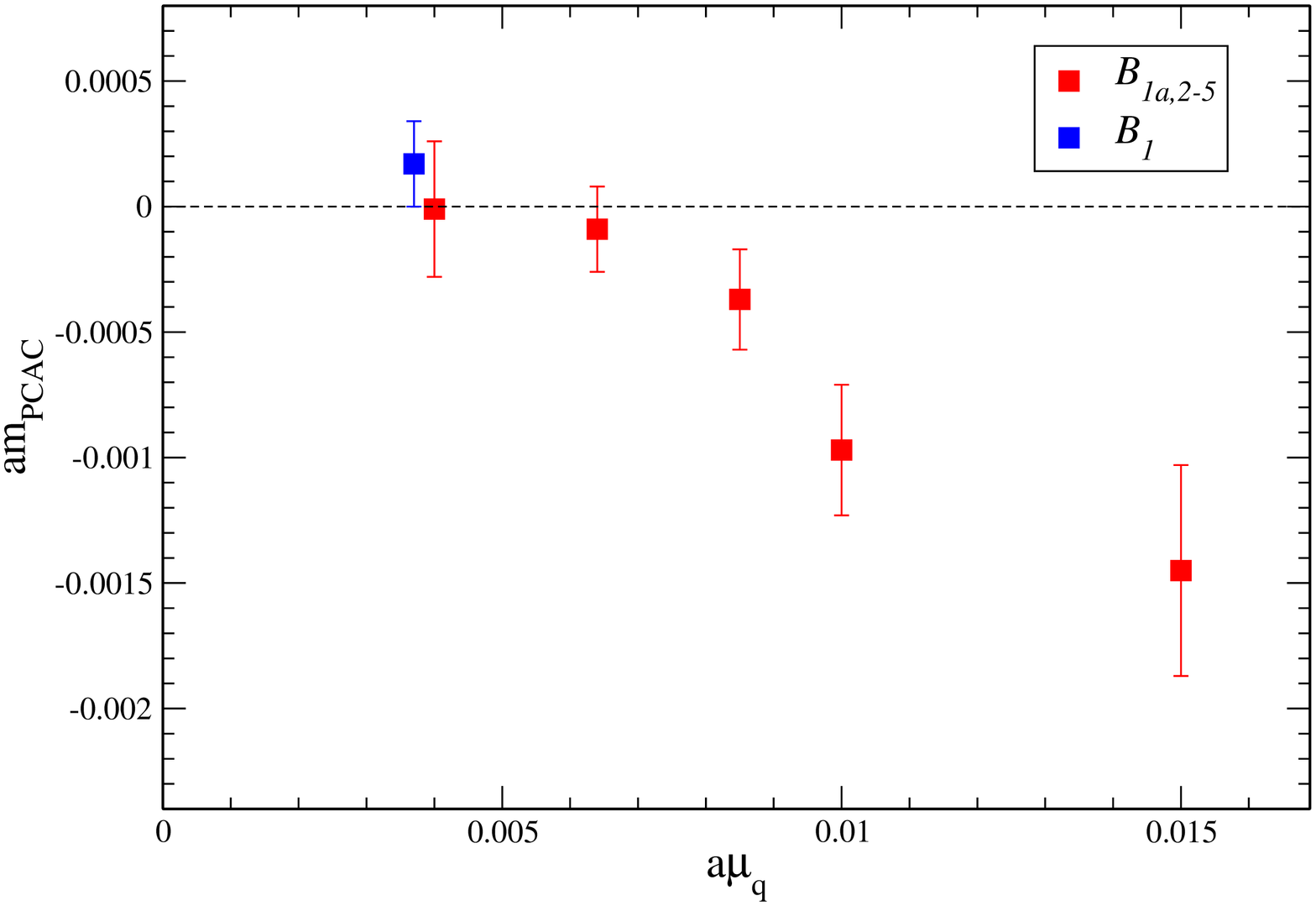}
  \caption{The PCAC quark mass $am_\mathrm{PCAC}$ as a function of $a\mu_q$}
    \label{fig:mpcac}
\end{figure}

Compared to ref.~\cite{Boucaud:2007uk} we now have a result for
$am_\mathrm{PCAC}$ available for the large statistics run $B_1$. It is 
reassuring that there is full consistency between the $5000$ trajectory
run and the run extended up to $10000$ trajectories. This makes us
confident that our error estimate is realistic. Our results for
$am_\mathrm{PCAC}$ as a function of the bare quark mass $a\mu_q$ are
illustrated in fig.~\ref{fig:mpcac}, where results from the full
ensemble, $B_1$, and those from the smaller ensemble, $B_{1a}$, are
separately shown.

\subsection{Pseudoscalar decay constant and $Z_{\rm V}$ }
\label{ssec:decay}

Using the exact lattice (twisted basis) PCVC relation~\footnote{We recall
that at
maximal twist the twisted basis vector current $V_\mu^a$ corresponds
to the axial current $\epsilon^{3ba}A'^b_\mu$ ($a,b = 1,2$) in the
physical quark basis.}
\begin{equation}
  \langle\partial^*_\mu \tilde V^a_\mu(x) O(0)\rangle= -2 \mu_{\rm q}
\epsilon^{3ab}
  \langle P^b(x) O(0)\rangle \qquad a=1,2 \, ,
  \label{PCVC}
\end{equation}
where $\partial^*_\mu$ is the lattice backward derivative,
$O$ a local lattice operator and $\tilde V^a_\mu(x)$ the 1-point-split
vector current
\begin{eqnarray}
  \label{eq:psplit_cur}
  \tilde{V}^a_{\mu}(x) =
  \frac{1}{4}\left [\overline{\chi}(x)\tau_a
U_{\mu}(x)(\gamma_\mu-r)\chi(x+\hat\mu)\ + \right.\nonumber \\
  \quad\quad\left. \overline{\chi}(x+\hat\mu)\tau_a
U^{\dagger}_{\mu}(x)(\gamma_\mu+r)\chi_{\mu}(x)\right ]\, ,
\end{eqnarray}
we can also compute (in the charged meson channel) the
pseudoscalar meson decay constant with no need of
any renormalization constant
(see~\cite{Frezzotti:2000nk,Frezzotti:2001du,Jansen:2003ir})
from the formula
\begin{equation}
  f_{\rm PS}=\frac{2\mu_{\rm q}}{m_{\rm PS}^2} | \langle
  0 | P^a | {PS}\rangle | \qquad a=1,2 \, .
  \label{indirect}
\end{equation}
Based on the exact relation~(\ref{PCVC}) and noting that
\begin{equation}
\partial^*_\mu \tilde{V}_\mu^a(x) \; = \; Z_{\rm V} \widetilde{\partial}_\mu
V_\mu^a(x)
+  {\rm O}(a^2) \, ,
\label{eq:Vc2V}
\end{equation}
where $V_\mu^a$ is the naive current defined in eq.~(\ref{operators})
and $\widetilde{\partial}_\mu $ the symmetric lattice derivative,
we can determine the (scale-independent) renormalization constant $Z_{\rm V}$
through ($a,b = 1,2$ and $x_0 \neq 0$)
\begin{equation}
  \label{eq:ZVbest}
  Z_\mathrm{V}(a\mu_q) = \frac{-2\mu_q \sum_{\bf x} \epsilon^{3ab} \langle P^b(x) P^b(0)
\rangle}
  {\sum_{\bf x} \widetilde{\partial}_0\langle V^a_0(x) P^b(0)
\rangle}\qquad\textrm{and}\qquad
  Z_\mathrm{V}=\lim_{\mu_q\rightarrow 0} Z_\mathrm{V}(a\mu_q)
\end{equation}
with only $\mathrm{O}(a^2)$ cutoff effects. 
The results for $a f_\mathrm{PS}$ extracted using eq.~(\ref{indirect})
are collected in Table~\ref{tab:resmpsfps} and will be discussed
further in the next section. 

In the same Table we quote the results
for the determination of $Z_\mathrm{V}$ that is based on fitting the 
large time behaviour of eq.~(\ref{eq:ZVbest}) where the pion state dominates 
\begin{equation}
  \label{eq:ZVlive}
\frac{-2\mu_q \sum_{\bf x} \epsilon^{3ab} \langle P^b(x) P^b(0)
\rangle}
  {\sum_{\bf x} \widetilde{\partial}_0\langle V^a_0(x) P^b(0) \rangle}
\quad \stackrel{x_0\;\;{\rm large}}{\rightarrow} \quad
\frac{ 2\mu_q \epsilon^{3ab} \langle \Omega | P^b | \pi^b({\bf 0}) \rangle }
{ m_{\rm PS} \langle \Omega | V^a | \pi^b({\bf 0}) \rangle } \, .
\end{equation}
In practice the relevant matrix elements are extracted from a 
factorising fit (see eq.~(\ref{eq:factorisingfit})) to a ($4 \times 4$) matrix
of correlators, with entries given by the expectation values of 
$\sum_{\bf x} P^1(x) P^1(0)$ and 
$\sum_{\bf x} V^2_0(x) P^1(0)$
with local-local and local-fuzzed operators. 

\begin{figure}[t]
  \centering
  \includegraphics[width=0.75\linewidth,angle=-90]{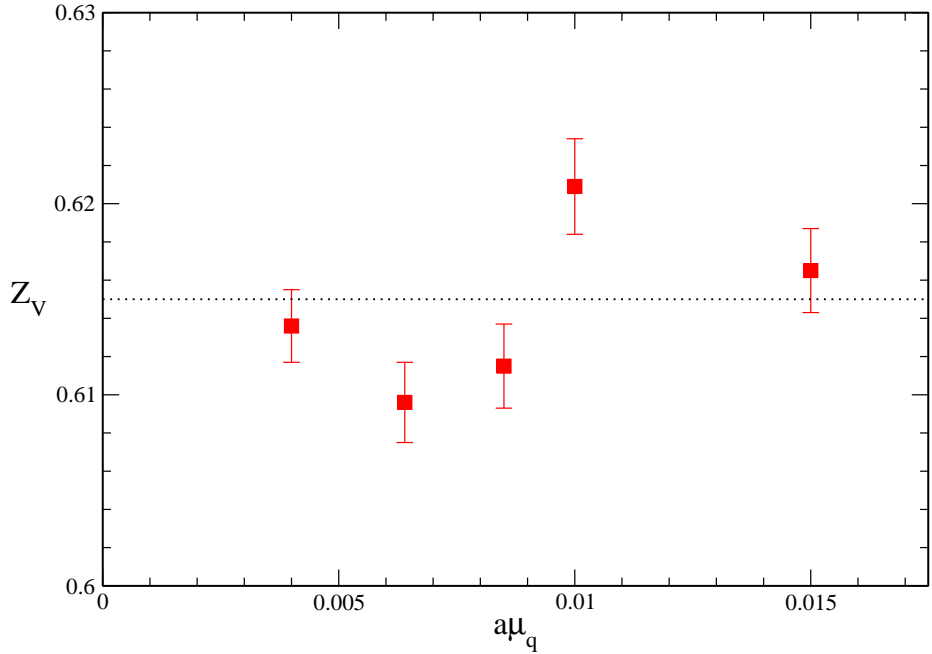}
  \caption{Extrapolation of $Z_{\rm V}$ to zero quark mass at
    $\beta=3.9$. The data are consistent within errors with the
 constant behaviour shown in the figure.}
    \label{fig:zv}
\end{figure}

The $a\mu_q$ dependence of $Z_{\rm V}$ is everywhere very weak 
(see fig.~\ref{fig:zv}),  
which makes the extrapolation to the chiral point easy and
almost irrelevant,
and leads to a rather precise result: $Z_{\rm V} = 0.615(5)$.
The quoted error is a conservative estimate of the total uncertainty
on $Z_{\rm V}$, inferred from the data in the last column of     
Table~\ref{tab:resmpsfps} and their statistical errors. 

Another determination of $Z_{\rm V}(a\mu_q)$ is obtained by evaluating the ratios of correlators in
the r.h.s.\ of eq.~(37) for a number of time separations ($x_0/a \geq 10$) and taking the average over
them.
This alternative method to evaluate $Z_{\rm V}$ has been presented in
ref.~\cite{Dimopoulos:2007fn}.
Applying it, for instance, to the available ($\sim$~900) 
correlators computed on the whole ensemble $B_1$ yields the result
$Z_{\rm V}(a\mu_q=0.0040)=0.6101(2)$. Both approaches, the one discussed
in some detail above and the one of ref.~\cite{Dimopoulos:2007fn}, 
provide precise determinations 
for $Z_{\rm V}$ and the virtues of both methods 
will be further discussed in ref.~\cite{ETMC:wip}.

\section{Chiral Perturbation Theory analysis of $f_\mathrm{PS}$ and $m_\mathrm{PS}$}  
\label{sec:xpt}  
  
In this section we present the details of the comparison of our data  
with Chiral Perturbation Theory ($\chi$PT) predictions. 
The main results have been already  
published in ref.~\cite{Boucaud:2007uk}. Here we provide further  
information about our fitting procedure and error analysis.  
The main goal of this section is to explain the fitting procedure and 
error determination using chiral perturbation theory. We will therefore 
only use the ensembles $B_1$-$B_5$, i.e. the ensembles that have already 
been discussed in ref.~\cite{Boucaud:2007uk}. For this limited set of data at
only one value of the lattice spacing of $a=0.087$fm, we are not sensitive to higher
order effects of chiral perturbation theory. We will therefore restrict 
ourselves to a 1-loop analysis of the data on $M_{\rm PS}$ and $f_{\rm PS}$. 
Nevertheless, we use the 2-loop chiral perturbation theory expressions and vary 
parameters of the corresponding formulae to see the possible effects, if the 
2-loop order would be important to describe our data.
In particular, this 2-loop investigation confirms that the here chosen 
dataset is indeed not sensitive to higher loop corrections.
 
Some preliminary results at larger volumes ($L/a=32$) and finer lattice spacing ($\beta=4.05$)  
have been already presented~\cite{Urbach:2007rt,Dimopoulos:2007qy}. 
However, the present work is focused on the  
details of the analysis of the data points presented in ref.~\cite{Boucaud:2007uk}. 
The study of the volume and scaling dependence will be presented elsewhere. 
Notice, however, that, w.r.t.\ ref.~\cite{Boucaud:2007uk} we have a larger statistics 
at the smallest quark mass. 
 
Our raw data for $am_\mathrm{PS}$ and $af_\mathrm{PS}$ are determined  
as described in sect.~\ref{sec:errors}. Results are reported in  
Table~\ref{tab:resmpsfps}.  As said above, there is no need to compute any  
renormalisation constant in order to make contact with the corresponding   
physical quantities.  
  
In our $\chi$PT based analysis we have to take into account finite size  
corrections because on our lattices at the lowest and next-to-lowest  
$\mu_q$-values they turn out to affect $am_\mathrm{PS}$ and, especially,   
$af_\mathrm{PS}$ in a significant way.   
  
We have used continuum  $\chi$PT to describe consistently the dependence of the data  
both on the  finite spatial size ($L$) and on the bare quark mass ($\mu_q$). This might 
look inappropriate in view of the existence of a large additive O($a^2$) artifact in the  
neutral pion mass 
squared~\footnote{Theoretical arguments have been presented~\cite{Frezzotti:2007qv} 
suggesting that this lattice artifact is an exceptional, though important, case, because 
it is related to the large value of a continuum matrix element appearing in the Symanzik 
expansion of the $\pi^0$-mass and does not stem from large coefficients multiplying the 
dimension five and six terms of the Symanzik effective Lagrangian.}. 
However this is not so, because theoretical analyses carried out in the framework 
of lattice $\chi$PT~\cite{Sharpe:2004ny} and Symanzik expansion (complemented 
with soft pion theorems in the continuum theory)~\cite{Frezzotti:2005gi,Frezzotti:2007qv} show that, if  
maximal twist is implemented as discussed in sect.~\ref{subsec:systematic}, 
the charged pion squared mass differs from its continuum counterpart only by
O($a^2\mu$) and O($a^4$) terms, while the charged pion decay constant is affected
by (chirally non-enhanced) discretization errors of order $a^2$. 
Moreover the Symanzik expansion analysis is 
applicable for all spatial volumes $L^3$, provided $L$ is large enough to justify the 
use of soft pion theorems in the continuum theory at the quark mass of interest. This 
entails that also the $L$-dependence of the charged pion squared mass and decay constant 
is expected to be essentially continuum-like. These expectations are also supported by  
preliminary and still partial results we obtain at different lattice resolutions and 
different physical volumes~\cite{Urbach:2007rt,Dimopoulos:2007qy}. Last but not least, the  
continuum $\chi$PT formulae appear to describe well our data, as we are going to show below.
 
We fit the appropriate ($N_f=2$) next-to-leading-order (NLO) $\chi$PT  
formulae~\cite{Gasser:1986vb,Colangelo:2005gd}  
\begin{equation}  
  m_\mathrm{PS}^2(L) = 2B_0\mu_q \, \left[  
    1 + \frac{1}{2}\xi \tilde{g}_1( \lambda )    \right]^{2} \, \left[ 1 +  
    \xi \log ( 2B_0\mu_q/\Lambda_3^2 ) \right] \, ,  
  \label{eq:chirfo1}  
\end{equation}  
\begin{equation}  
  f_\mathrm{PS}(L) = F \, \left[  
    1 - 2 \xi \tilde{g}_1( \lambda )    \right] \, \left[ 1 -  
    2 \xi \log ( 2B_0\mu_q/\Lambda_4^2 ) \right] \, ,  
  \label{eq:chirfo2}  
\end{equation}  
to   
our raw data for $m_\mathrm{PS}$ and $f_\mathrm{PS}$  
simultaneously. Here~\footnote{We stress that $\xi$ defined here   
should not be confused with the continuum matrix element $\xi_\pi$  
introduced in eq.~(\ref{eq:xipi}).}  
\begin{equation}  
  \xi = 2B_0\mu_q/(4\pi F)^2 \, , \qquad  
  \lambda = \sqrt{2B_0\mu_q L^2}\, .    
\end{equation}  
The finite size correction function $\tilde{g}_1(\lambda)$ was first  
computed by Gasser and Leutwyler in ref.~\cite{Gasser:1986vb} and is  
also discussed in ref.~\cite{Colangelo:2005gd} from which we take our  
notation (except that our normalisation of $f_{\pi}$ is 130.7 MeV).   
In eqs.~(\ref{eq:chirfo1}) and~(\ref{eq:chirfo2}) next-to-next-to-leading order  
(NNLO) $\chi$PT corrections are assumed to be negligible   
(this assumption is critically discussed in sect.~\ref{subsec:staterrxpt}).   
The formulae above depend on four unknown parameters,   
$B_0$, $F$, $\Lambda_3$ and $\Lambda_4$, which will be determined by   
fitting to our data.  
  
\subsection{Statistical errors}  
\label{subsec:staterrxpt}

In order to estimate the errors on the   
fit parameters it is  
important to account for both autocorrelation and cross-correlation of  
the data. We have exploited two different methods to do so.  
  
\vspace{0.3cm}  
\noindent\emph{Method A} \\ 
\noindent The first method 
(see also~\cite{Farchioni:2005bh}) 
consists in computing the full covariance matrix of our data  
for $af_\mathrm{PS}$ and $(am_\mathrm{PS})^2$ and include it in the  
computation of $\chi^2$  
\begin{equation}  
\chi^2 = \sum_{i,j} (y_i-F_i) V^{-1}_{i,j} (y_j-F_j) \, ,
\label{eq:chi2cov} 
\end{equation}  
where $V$ is the covariance matrix~\footnote{As we have data from independent  
simulations (ensembles $B_1$ to $B_5$), in the present case   
the covariance matrix will be block-diagonal with five blocks.}  
\begin{equation}  
V_{i,j} = {\rm cov}(y_i,y_j) = {\rm cov}((y_i-F_i)(y_j-F_j)) \, ,   
\label{eq:covmat} 
\end{equation}  
normalised so that the diagonal elements coincide with the squared standard  
error, and we have expressed the $\chi$PT ansatz, eqs.~(\ref{eq:chirfo1})  
and~(\ref{eq:chirfo2}), in the form  
\begin{equation}  
  \label{eq:norm_form}  
  y_i=F_i(\vec{x},\vec{\theta})\, .  
\end{equation}  
Here we denote by $y_i$ the primary measured quantities (in this  
section: $y_1=(am_{\rm PS}|_{B_1})^2$, $y_2=af_{\rm PS}|_{B_1}$,   
$y_3=(am_{\rm PS}|_{B_2})^2$, $\dots$), by $\vec{x}$ the independent   
(error-free) variables (in this section: only $x= a\mu_q$) and by  
$\vec{\theta}$ the parameters to be determined by the fit (here:  
$\theta_1=2aB_0$, $\theta_2=aF$, $\theta_{3,4} = \log(a^2\Lambda_{3,4}^2))$.  
The error on the parameters are thus given by  
\begin{equation}  
(\Delta \theta_\alpha)^2= (\nabla_{\theta_\alpha}  
F^T V^{-1} \nabla_{\theta_\alpha} F)^{-1} \, .  
\label{eq:paramerr} 
\end{equation}  
  
The autocorrelations of $(am_\mathrm{PS})^2$ and $af_\mathrm{PS}$ have  
been estimated both by data-blocking and by means of the  
$\Gamma$-method, as discussed in sect.~\ref{sec:errors}.   
Both approaches indicate (see Table~\ref{tab:autocorr})  
that by combining data into blocks 
of 32 measurements each (this corresponds always to more 
than 60 MC trajectories) the resulting blocked data are 
safely uncorrelated. These blocked data are thus used 
to evaluate the covariance matrix, the $\chi^2$ and the 
errors on the fit parameters as discussed above  
(eqs.~\ref{eq:chi2cov}--\ref{eq:paramerr}). 
In this way the possible effect of cross-correlations among  
the observables is included in the covariance matrix and therefore  
properly accounted for in the fit procedure.  
  
In some of the checks that we are going to present below it will not  
always be possible to reduce the $\chi$PT formulae to the form of  
eq.~(\ref{eq:norm_form}). This happens in particular in the following cases: when computing  
directly $af_\mathrm{PS}$ as a function of $am_\mathrm{PS}$, when  
including the effects due to a non-vanishing $am_\mathrm{PCAC}$,  
when including higher orders in Finite Size Effects (FSE) calculations, as computed  
in~\cite{Colangelo:2005gd}, or when we will eventually   
study the scaling dependence for different lattice spacings~$a$. In  
all these cases the $\chi$PT formulae can be expressed in the more general form  
$G_i(\vec{y},\vec{x},\vec{\theta})=0$ and the errors   
are given by the formula~\cite{BrittLuecke}: $(\Delta \theta_\alpha)^2= (\nabla_{\theta_\alpha}  
G^T (\nabla_y G V \nabla_y G^T)^{-1} \nabla_{\theta_\alpha} G)^{-1}$.  
These errors are obtained as propagation from the known  
errors on $\vec{x}$, therefore they do not depend on how good the fit is.   
The quality of the fit is expressed as usual by the quantity $\chi^2/{\rm d.o.f.}$.  
  
To provide a further check of possible effects of cross-correlation,  
the fits were also performed by dividing the data set into two  
subgroups each of half the size. The data for $am_\mathrm{PS}$ were taken from  
one subgroup of gauge configurations and those for $af_\mathrm{PS}$  
from the other, ensuring in this way absence of cross-correlation.   
Errors scale as $\sqrt{2}$, i.e.\ as expected from halving  
the statistic, which indicates a negligible effect of  
cross-correlations in the full data set. Stability was also checked  
against different choices of subgroups. This result is confirmed by  
the observation that if we suppress the off-diagonal terms in the  
covariance matrix, our error bars are affected only at the percent  
level.  
  
\vspace{0.3cm}  
\noindent\emph{Method B}  \\ 
\noindent\emph{Method A} is standard, and of course unbiased if  
for all observables the data are distributed   
in a Gaussian way (which we checked explicitly to be the  
case to a good approximation)  
and if the functions $F$  
(or $G$) have a sufficiently linear behaviour   
around the relevant values of their  
arguments. An even safer   
estimate of the final errors can be obtained with  
the bootstrap method~\cite{EfronTibshirani:1993,davison97}.  
  
To apply the bootstrap analysis method to our data set we proceed as follows.   
In order to account for autocorrelations we first   
form bins of 32 gauge configurations for each value of   
$\mu_q$. Out of the blocked data we  
generate 1000 bootstrap samples. The size of each sample is chosen as  
large as the full (blocked) data set. From the 1000 bootstrap samples  
we obtained 1000 \emph{observations} for $2aB_0$, $aF$,  
$\log(a^2\Lambda_{3,4}^2)$, and $\bar{l}_{3,4}\equiv \log(\Lambda_{3,4}^2/m_{\pi}^2)$, 
respectively. Error  
estimates are then computed as prescribed by the bootstrap method,  
i.e.\ by the standard deviation   
over the (equally weighted) 1000 samples.    
Incidentally we remark that this procedure takes the cross correlation  
between $am_\mathrm{PS}^2$ and $af_\mathrm{PS}$ correctly into account.  
In the 1000 fits we performed we have hence always used only the diagonal  
elements of the covariance matrix (fixed to their   
central values, i.e.\ to the square of the 
statistical errors on $am_\mathrm{PS}^2$ and $af_\mathrm{PS}$, 
see Table~\ref{tab:resmpsfps})
as weights to evaluate the $\chi^2$ 
formula~(\ref{eq:chi2cov}).
In this specific application of the bootstrap method
the errorbars on the basic quantities $am_\mathrm{PS}^2$ and $af_\mathrm{PS}$ 
are still needed, since the observables
of interest, the low energy constants (LEC), 
are defined through minimization of the $\chi^2$
of the simultaneous fit to the 
$\chi$PT formulae eqs.~(\ref{eq:chirfo1})--(\ref{eq:chirfo2}).
Note that to safely employ the bootstrap method data need not have a Gaussian  
distribution, nor do the constraints, defined by the $\chi$PT   
formulae, need to be linear.   
The bootstrap method may become expensive if single fits are   
significantly computer time demanding.   
  
Both methods A and B give consistent results, as shown in  
Table~\ref{tab:ABCfits}.   
In this paper we use the same setup as in ref.~\cite{Boucaud:2007uk}, but we   
employ a somewhat larger   
statistics. The results are consistent.  
In addition to the error estimates we quote   
the value of $\chi^2$ and the   
merit figure of the fit defined via  
\[  
Q = 1 - P(\chi^2/2,\ \mathrm{d.o.f.}/2)\, ,  
\]  
where $P$ is the incomplete Gamma function~\cite{NumRec}. 
 
\begin{table}[t]  
  \centering  
  \begin{tabular*}{.7\linewidth}{@{\extracolsep{\fill}}rccc}  
    \hline\hline  
      $\Bigl.\Bigr.$ & Method A & Method B \\  
    \hline\hline  
    $\bigl.\bigr.2aB_0$ & 5.04(7) & 5.04(7)  \\  
    $\bigl.\bigr.aF$ & 0.0522(7)  & 0.0522(7) \\  
    $\bigl.\bigr.\log(a^2\Lambda_3^2)$ & -1.90(11) & -1.91(10) \\  
    $\bigl.\bigr.\log(a^2\Lambda_4^2)$ & -1.00(4)  & -1.00(4)  \\  
    $\bigl.\bigr.\chi^2/{\rm d.o.f.} $ &  1.0/4     &  0.9/4  \\  
    $\bigl.\bigr.Q$                    &  0.91      &  0.92  \\  
 
    \hline\hline  
  \end{tabular*}  
  \caption{Comparison of fit results from methods A and B.}  
  \label{tab:ABCfits}  
\end{table}  
  
\subsection{Discussion of systematic errors}  
\label{subsec:systematic}
  
The error bars quoted in Table~\ref{tab:ABCfits}   
are only statistical. As we also stressed in  
ref.~\cite{Boucaud:2007uk}, a number of systematic effects are  
expected. Here we present some checks we performed in order to  
estimate the actual magnitude of these systematic effects.   
  
As a first, simple check on the possible impact of neglected NNLO terms on the results
presented in Table~\ref{tab:ABCfits}, we have also included
the heaviest point (the one at $a\mu_q=0.0150$) in the
standard fit to the formulae~(\ref{eq:chirfo1})--(\ref{eq:chirfo2}).
In this case the results are still compatible with those in Table~\ref{tab:ABCfits}
within 1.7 standard deviations, but the $\chi^2/{\rm d.o.f.}$ of the fit jumps
from $0.24$ to $1.7$. This increase of $\chi^2/{\rm d.o.f.}$ is mainly due
to the point at $af_{\rm PS}$ at $a\mu_q=0.0150$, as we noted 
already in ref.~\cite{Boucaud:2007uk}.
The results of the fit are displayed in the second column of
Table~\ref{tab:allfits}. This suggests that only the first four quark mass points
should be used when comparing our data for $af_{\rm PS}$ and $am_{\rm PS}$
with NLO $\chi$PT, as was done in ref.~\cite{Boucaud:2007uk}.
  
It is also very interesting to see how much the tiny deviations from  
maximal twist corresponding to the (statistically compatible  
with zero) measured central values of $m_{\rm PCAC}$ affect our results  
for the low energy constants discussed in this section.  
To address this question we introduce the definition of bare quark mass,   
$m_q=\sqrt{(Z_A m_\mathrm{PCAC})^2 + \mu_q^2}$, which holds for generic twist angle  
up to neglected O($am_{\rm PCAC}$) and O($a^2$) terms. Moreover, in order to  
take into account the axial-$\tau_3$ transformation properties of the current  
entering the formal definition of $f_{\rm PS}$, at the same level of accuracy,   
the value of $af_{\rm PS}$ should be corrected into $af_{\rm PS}m_q/\mu_q$.   
We remark   
that this is obtained automatically if $f_{\rm PS}$ is evaluated from eq.~(\ref{indirect})  
with $\mu_q$ replaced by $m_q$ -- this can be related to the invariance of the   
operator $P^{1,2}$, a matrix element of which appears on the r.h.s.\ of eq.~(\ref{indirect}), 
under axial-$\tau_3$ rotations. The results of this analysis, where  
we set $Z_A=0.76(2)$, as found at $\beta=3.9$ in ref.~\cite{Dimopoulos:2007fn},  
are shown in the last column of Table~\ref{tab:allfits}. 
It is reassuring to see that, thanks to  
the good precision we could reach in setting $m_{\rm PCAC}$ to zero, the  
low energy constants of interest here are left essentially unaffected by  
this kind of correction.   

\begin{table}[t]
  \centering
  \begin{tabular*}{1.\linewidth}{@{\extracolsep{\fill}}rcc}
    \hline\hline
     $\Bigl.\Bigr.$ & including $a\mu_q=0.015$ &
     $m_q=\sqrt{(Z_\mathrm{A} m_\mathrm{PCAC})^2 + \mu_q^2}$ \\
    \hline\hline
    $\bigl.\bigr.2aB_0$                &  5.06(5)   &  5.05(6) \\
    $\bigl.\bigr.aF$                   &  0.0508(5) &  0.0521(7) \\
    $\bigl.\bigr.\log(a^2\Lambda_3^2)$ & -1.93(6)   & -1.87(11) \\
    $\bigl.\bigr.\log(a^2\Lambda_4^2)$ & -0.89(2)   & -0.99(4)\\
    $\bigl.\bigr.\chi^2/{\rm d.o.f.} $ &  10.3/6    &  0.55/4 \\
    $\bigl.\bigr.Q$                    &  0.11      &  0.97\\
    \hline\hline
  \end{tabular*}
  \caption{Comparison of fit results from different setups, as
    explained in the text.}
  \label{tab:allfits}
\end{table}

We now consider the finite size corrections. In ref.~\cite{Boucaud:2007uk}  
we estimated them with the help of the formulae of ref.~\cite{Gasser:1986vb}. A  
nice feature of these formulae is that they introduce no new  
parameter. However, they are only the first term of an  
expansion. Hence, the question is: how large is the residual  
uncertainty in FSE due to this truncation? 
To go beyond the first term in the framework of ref.~\cite{Gasser:1986vb} 
is difficult. For the pseudo scalar mass the FSE corrections  
at two loops in $\chi$PT have been computed in ref.~\cite{Colangelo:2005gd}.   
However, one can do better using the kind of $\chi$PT expansion   
suggested in ref.~\cite{Colangelo:2003hf}, for  
which results are also given in ref.~\cite{Colangelo:2005gd}.   
  
With the help of the results from ref.~\cite{Colangelo:2005gd}  
we can assess the stability of the prediction both  
by comparing the two approaches and by    
studying the convergence of  
the expansion of refs.~\cite{Colangelo:2003hf,Colangelo:2005gd}.   
One should also notice that higher orders do  
introduce new parameters. Since it is not realistic to fit them,   
we will instead look at the stability of the prediction while changing  
those parameters in a ``reasonable'' range. The ``reasonable'' range  
is suggested in ref.~\cite{Colangelo:2005gd} and is based on  
phenomenological grounds.   
  
To avoid confusion, we remark that the  
results of ref.~\cite{Gasser:1986vb} are given as an expansion in  
powers of $1/F_0$, while ref.~\cite{Colangelo:2005gd} uses an  
expansion in $1/F_{\pi}$. This is the only reason why the first term  
of ref.~\cite{Colangelo:2005gd} does not coincide with  
ref.~\cite{Gasser:1986vb}.   
  
In Tables~\ref{tab:FSE} and~\ref{tab:FSEfps} we show the percent deviation obtained using the formulae 
from refs.~\cite{Gasser:1986vb} and~\cite{Colangelo:2005gd} at different orders. Note that the new  
low energy constants (LECs) that at higher orders of $\chi$PT are relevant  
for FSE are fixed to their central values estimated in    
ref.~\cite{Colangelo:2005gd}. See the comment below about their impact.  
To distinguish the expansion of the FSE effects from the usual $\chi$PT expansion  
we will use a lower case notation (lo, nlo, nnlo) to denote the former one.  
The two expansions are of course related, but since the FSE also   
depend on the lattice size $L$,  
there is no reason to truncate the chiral expansion for FSE at the same order as  
the usual $\chi$PT expansion. Here, for instance, we will  
use the NLO $\chi$PT formulae, but we will compare FSE at lo, nlo and nnlo.  
  
\begin{table}[t]  
  \centering  
  \begin{tabular*}{.85\linewidth}{@{\extracolsep{\fill}}rcccc}  
    \hline\hline  
     $a\mu_q$ & lo \cite{Gasser:1986vb} & lo \cite{Colangelo:2005gd} &  
     nlo \cite{Colangelo:2005gd} & nnlo \cite{Colangelo:2005gd} \\  
    \hline\hline  
    0.0040 & 0.64 \% & 0.42 \% &  0.50 \% &  0.21 \% \\  
  
    0.0064 & 0.29 \% & 0.16 \% &  0.21 \% &  0.10 \% \\  
  
    0.0085 & 0.16 \% & 0.08 \% &  0.12 \% &  0.06 \% \\  
  
    0.0100 & 0.11 \% & 0.05 \% &  0.08 \% &  0.04 \% \\  
    \hline\hline  
  \end{tabular*}  
  
  \caption{Percent Finite Size deviation  
    $(m_\mathrm{PS}(L)-m_\mathrm{PS}(\infty))/m_\mathrm{PS}(\infty)$  
    predicted by $\chi$PT for our data points. Note that nlo and nnlo  
    include only the last order and not the previous one(s). According to  
    ref.~\cite{Colangelo:2003hf}, comparing nlo and nnlo (not lo and nlo)
    gives a reliable indication about the convergence of the expansion.}
  \label{tab:FSE}  
\end{table}  
  
\begin{table}[t]  
  \centering  
  \begin{tabular*}{.7\linewidth}{@{\extracolsep{\fill}}rccc}  
    \hline\hline  
     $a\mu_q$ & lo \cite{Gasser:1986vb} & lo \cite{Colangelo:2005gd} &  
     nlo \cite{Colangelo:2005gd} \\  
    \hline\hline  
    0.0040 & -2.57 \% & -1.68 \% &  -0.76 \% \\  
  
    0.0064 & -1.15 \% & -0.63 \% &  -0.30 \% \\  
  
    0.0085 & -0.64 \% & -0.32 \% &  -0.16 \% \\  
  
    0.0100 & -0.44 \% & -0.21 \% &  -0.11 \% \\  
    \hline\hline  
  \end{tabular*}  
  \caption{Same as in Table~\ref{tab:FSE}, but for  
    $(f_\mathrm{PS}(L)-f_\mathrm{PS}(\infty))/f_\mathrm{PS}(\infty)$.}  
  \label{tab:FSEfps}  
\end{table}  
  
The convergence of the FSE expansion is expected to be good for all our  
data points since the smallest value of $m_\mathrm{PS} L$ is larger  
than 3. We recall that, according to ref.~\cite{Colangelo:2003hf}, the  
comparison of lo and nlo is not a good indicator of the convergence of  
the expansion. This should be rather checked by comparing nlo and  
nnlo. According to all our estimates only the FSE at the lightest point  
($a\mu_q=0.004$) are relevant, while those at larger quark masses   
are always smaller than statistical errors. For instance,   
the deviations in $m_\mathrm{PS}$ are barely larger than its statistical errors (which  
amount to about $0.5\%$).   
In order to check   
the dependence of the predicted FSE corrections on the LECs entering   
only at nlo,    
we changed randomly the value of the latter within the ``reasonable'' range suggested in  
ref.~\cite{Colangelo:2005gd}. We saw that nlo and nnlo FSE corrections are affected    
only at the level of about 20\% (lo corrections are obviously unaffected) by such changes.

Up to this point we have only considered the $\chi$PT at NLO   
(however corrections as high as nnlo are included in FSE calculations)   
implicitly assuming that NNLO contributions are  
negligible. This is reasonable, since $\chi$PT  formulae   
with only NLO corrections yield a very good fit of the data at the   
lightest four quark masses, in spite of the fact that the expansion parameter,  
$\xi = 2B_0 \mu_q / (4 \pi F)^2$, 
is not always very small.  
It is thus important to assess how much NNLO terms would  
affect our results.  
  
The NNLO corrections relevant for $m_\mathrm{PS}$ and $f_\mathrm{PS}$  have been  
calculated in ref.~\cite{Bijnens:1997vq}. Here we use an  expression  
which is easier to compare with lattice data, namely the one of   
refs.~\cite{Leutwyler:2000hx,Col:comm} which reads   
\begin{equation}  
  \begin{split}  
    m_\mathrm{PS}^2 &= M^2 \left\{ 1 + \xi\log{\frac{M^2}{\Lambda_3^2}} +   
      \frac{17}{2} \xi^2 \left[\log{\frac{M^2}{\Lambda_M^2}}\right]^2   
      + 4 \xi^2 k_M + {\rm O}(\xi^3) \right\}\; , \\  
    f_\mathrm{PS} &= F\ \left\{ 1 - 2 \xi \log{\frac{M^2}{\Lambda_4^2}} -  
      5 \xi^2 \left[\log{\frac{M^2}{\Lambda_F^2}}\right]^2 + 4 \xi^2 k_F + {\rm O}(\xi^3)  
    \right\}  \, , 
  \end{split}  
\label{eq:NNLO}  
\end{equation}  
where $\xi=2B_0\mu_q/(4 \pi F)^2$ as before, $M^2=2 B_0\mu_q$ and  
\begin{eqnarray*}  
\log{ \frac{\Lambda_M^2}{M^2} } &=& \frac{1}{51} \left(   
28 \log{ \frac{\Lambda_1^2}{M^2} } +   
32 \log{ \frac{\Lambda_2^2}{M^2} } -  
9 \log{ \frac{\Lambda_3^2}{M^2} } +   
49 \right)\; , \\  
\log{ \frac{\Lambda_F^2}{M^2} } &=& \frac{1}{30} \left(   
14 \log{ \frac{\Lambda_1^2}{M^2} } +   
16 \log{ \frac{\Lambda_2^2}{M^2} } +  
6 \log{ \frac{\Lambda_3^2}{M^2} } -  
6 \log{ \frac{\Lambda_4^2}{M^2} } +   
23 \right)\, . \\  
\end{eqnarray*}  
  
\begin{table}[t]  
  \centering  
   \begin{tabular*}{1.05\linewidth}{@{\extracolsep{\fill}}rccccc}  
    \hline\hline  
 & NNLO as in \cite{Colangelo:2005gd}  
 & $\delta \Lambda_1=\pm 33\%$ & $\delta \Lambda_2=\pm 5\%$ & $k_M=\pm 1$ & $k_F=\pm 1$ \\  
    \hline\hline  
    $2aB_0$ & 4.80(6)   & -0.66\% & -0.20\%  & 3.2\%  & 0.07\% \\  
            &           &  3.44\% &  0.26\%  &-2.5\%  &-0.12\% \\  
    \hline  
    $aF$    & 0.0536(6) & 0.60\%  &  0.16\%  &-0.19\% & 1.9\%  \\  
            &           & -1.7\%  & -0.19\%  & 0.21\% &-2.1\%  \\  
    \hline  
    $\log(a^2\Lambda_3^2)$ & -2.13(12) & -9.6\% & -1.2\%  & -29\% & -1.3\% \\  
                           &           & -5.9\% &  0.87\% &  26\% &  1.5\% \\  
  
    \hline  
    $\log(a^2\Lambda_4^2)$ & -1.00(5) & -4.6\%  & -0.50\% & 1.3\% &  24\% \\  
                           &          & -0.35\% &  0.34\% &-1.3\% & -26\% \\  
  
    \hline  
    $\chi^2/{\rm d.o.f.} $ & 0.085    &    1.7  &  1.1    &  0.48 &  1.4 \\  
                           &          &   0.15  &  0.82   &   1.8 &  0.73 \\  
  
    \hline\hline  
  \end{tabular*}  
  \caption{Fit results, including NNLO $\chi$PT. The second column shows the results obtained  
with the choice of $\Lambda_{1,2}$ suggested in~\cite{Colangelo:2005gd} and $k_{M,F}=0$.   
The other columns give the percent correction due to changing the corresponding parameter  
in the indicated range. For each line, the upper (lower) number corresponds to the higher (smaller)   
boundary value of the interval.}  
  \label{tab:NNLO}  
\end{table}  
 
It is not realistic to attempt a fit of all the coefficients involved in the full  
NNLO expressions at least with the limited set of data used here. 
Rather we fix the parameters $\Lambda_1$, $\Lambda_2$, $k_F$ and $k_M$ to the values  
suggested in ref.~\cite{Colangelo:2005gd}. Since no estimate for $k_{M,F}$ is available,   
we take $k_{M,F}=0$. Redoing the fit in these conditions we can check how much NNLO terms   
change the results of Table~\ref{tab:ABCfits}.   
The new fit results are shown in the second column of  
Table~\ref{tab:NNLO}. In order to further estimate to which extent these numbers are  
sensitive to a change in the parameters which were held fixed,   
we decided to change them one by one within the range proposed in  
refs.~\cite{Colangelo:2005gd,Colangelo:2003hf}, and perform a new fit for each one of these values.   
As for $k_{M}$ and $k_{F}$, it is difficult to tell what is a reasonable range,  
since, as we said, no estimate is available for them. On general grounds the values   
of $k_{M,F}$ are expected to be  
of O(1) and somewhat arbitrarily we assume a variability range $k_{M,F}=\pm 1$.  
This choice is also justified by the fact that larger variations quickly lead to very bad $\chi^2$.  
The results of this elaborated procedure are shown in columns 3 to 6 of Table~\ref{tab:NNLO}.  
Most effects are not significant if compared to statistical errors, as they are never larger   
than a few standard deviations. It should be noted, however, that $\Lambda_3$ appears to be rather   
sensitive to $k_M$ and similarly $\Lambda_4$ to $k_F$.  
These LECs can deviate by about 25\% when  setting $k_{M,F}$ to $+1$ or to $-1$.  
We mention that changes of the LECs similar to those reported in Table~\ref{tab:NNLO}  
are also obtained if the NNLO terms in eq.~(\ref{eq:NNLO}) are replaced by simple polynomial  
terms, like $\rho_{M,F} \xi^2$  (with no logarithms), and the free parameters $\rho_{M,F}$ are set  
to their best fit values.  
 
\subsection{Comments}  
  
In summary the discussion developed in this section shows that at least the systematic
errors coming from the unknown NNLO terms 
involving $k_{M,F}$ may be significantly larger than the
statistical ones, mostly because the adopted range of values was, to some degree,
arbitrarily chosen. 
However, as already said above, 
using only the datasets $B_1$-$B_5$ a reliable estimate of systematic 
uncertainties on $B$, $F$, $\Lambda_3$ and $\Lambda_4$
from the NNLO corrections is not possible.
A better assessment about the magnitude of NNLO effects
will be attempted elsewhere~\cite{ETMC_scaling} using ETMC
data at different lattice spacings.

Although FSE to our simulation data turn out to be less than a few percent, 
we have made a special effort to compute them quite accurately, because their impact 
on LECs cannot be neglected, as their magnitude is comparable to the size of our 
statistical errors.
The computation of FSE made in ref.~\cite{Colangelo:2005gd} represents a   
considerable improvement on the classical estimate of ref.~\cite{Gasser:1986vb}, 
as uncertainties on the extra LECs entering the former computation at high orders have little 
impact on the results.   
Actually, the validity of the predictions of FSE from $\chi$PT can be checked   
by performing simulations on lattices of increasing size in physical units.   
Preliminary results have been presented in ref.~\cite{Urbach:2007rt}.

\section{Summary} 
\label{sec:summary}

In this paper we have illustrated and discussed a number of details concerning 
unquenched simulations of $N_f=2$ mass degenerate Wilson quarks at maximal twist. 
We have explained in sect.~\ref{sec:twismasfer} 
our criterion on how to tune the theory to maximal twist. In particular, we provided  
theoretical arguments for our choice of $m_\mathrm{PCAC}/\mu_q\le 0.1$ and  
showed that an error $\Delta m_\mathrm{PCAC}/\mu_q \le 0.1$ is appropriate for this purpose.  
Useful formulae for quark bilinears and their physical interpretation 
in different quark bases (twisted and physical) are collected in Appendix~\ref{sec:appendixA}.
 
We have then discussed in sect.~\ref{sec:chargedmesons}  
the methods we have used to compute charged meson correlators  
emphasizing the effectiveness of employing (fuzzed) stochastic  
time-slice sources in the so-called ``one-end trick''.  
We have demonstrated that this method complemented by a  
random choice of the source location 
leads to a significant noise reduction, at least   
for two-point correlators in the meson sector.  
 
The computation of neutral mesons and, in particular,  
quark-disconnected 
contributions has been described in sect.~\ref{sec:neutralmesons} 
and in the corresponding Appendix~\ref{sec:appendixB}.  
We have spelled out the reasons for using stochastic volume sources which 
can be employed in combination with efficient variance reduction methods. 
All these technical improvements have allowed us to compute  
quark-disconnected contributions on our sets of  
unquenched 
gauge configurations to an acceptable 
accuracy. 
 
In sect.~\ref{sec:algo} we have illustrated the main features of 
the MC algorithms used in our simulations 
showing that the resulting 
autocorrelation times 
are small enough to allow for a trustworthy error analysis of physical 
observables. We also explain how our error analysis of the data 
was performed owing to the use of $\Gamma$- and binning-methods.  
 
The force parameter $r_0$ can serve as an important physical quantity to 
check the scaling behaviour towards the continuum limit. 
We have provided in sect.~\ref{sec:scale_static_potential} a comprehensive 
discussion of the methods we have used to extract $r_0$ on our 
configurations. It turns out that with the present data an accuracy of better 
than 1\% can be reached for $r_0$ in the chiral limit. It is also found that 
$r_0$ has a mild quark mass dependence which is consistent with being quadratic in $\mu_q$. 
 
Various results for the charged and neutral pseudoscalar masses, the  
untwisted PCAC quark mass and the renormalization constant $Z_{\rm V}$ are 
collected in sect.~\ref{sec:results}. In particular, we show  
``effective mass'' plots demonstrating 
the stability of the Euclidean time plateaux, 
which enables us to extract precise results for mesonic quantities. 
 
Finally, we have detailed in sect.~\ref{sec:xpt} how our $\chi$PT analysis of 
the data on $m_\mathrm{PS}$ and $f_\mathrm{PS}$ has been carried out, explaining 
how we get errors on the  
fitted low energy constants of the effective chiral Lagrangian, $B_0$, $F$, 
$\Lambda_3$ and $\Lambda_4$. In addition, we have analyzed the effects 
of higher orders in $\chi$PT on the stability of fit parameters and discussed 
the finite size effects. 
 
We consider the present paper as a technical reference work   
of our collaboration. The methods described  
here have been and will be extensively used in our ongoing future  
research on lattice QCD employing maximally  
twisted  
Wilson fermions.

\subsubsection*{Acknowledgments} 
 
We thank all other members of the ETMC for 
very valuable discussions and for a most enjoyable and 
fruitful collaboration.  
We also gratefully acknowledge discussions with D.~Be\'cirevi\'c   
and N.~Christian. 
The  computer time for this project was made available to us by the 
John von Neumann-Institute for Computing on the JUMP and Jubl systems 
in J\"ulich and apeNEXT system in Zeuthen, by UKQCD 
on the QCDOC machine at Edinburgh, by INFN and CNRS on the apeNEXT systems in Rome,  
by BSC on MareNostrum in Barcelona (www.bsc.es) and by the Leibniz Computer 
centre in Munich on the Altix system. We thank these computer centres and 
their staff for technical advice and help.  
On QCDOC we have made use of Chroma~\cite{Edwards:2004sx} and BAGEL~\cite{BAGEL} 
software and we thank members of UKQCD for assistance. For the analysis we used 
among others the R language for statistical computing~\cite{R:2005}. 
 
This work has been supported in part by  the DFG  
Sonder\-for\-schungs\-be\-reich/Transregio SFB/TR9-03, DFG project
JA 674/5-1 and the EU Integrated 
Infrastructure Initiative Hadron Physics (I3HP) under contract 
RII3-CT-2004-506078.  We also thank the DEISA Consortium (co-funded by 
the EU, FP6 project 508830), for support within the DEISA Extreme 
Computing Initiative (www.deisa.org).  G.C.R. and R.F. thank MIUR (Italy) 
for partial financial support under the contracts PRIN04 and and PRIN06.  
V.G. and D.P. thank MEC (Spain) for partial financial support under grant  
FPA2005-00711. 
 
\addappheadtotoc 
\appendixpage 
\appendix 
\section{Quark bilinear operators in the twisted basis}
\label{sec:appendixA}

We give in this appendix the expression of a number of bare
quark bilinear operators that are relevant for the topics of this 
paper. The operators are expressed in terms of 
i) simple composite fields
(recall $\gamma_5=\gamma_0\gamma_1\gamma_2\gamma_3$
and $\sigma_{\mu\nu} = i/2[\gamma_\mu,\gamma_\nu]$)
in the twisted quark basis,
where the fermionic action takes the form~(\ref{eq:Sf}), 
\begin{equation}
  \begin{split}
    S^0(x) & =  \bar{\chi}(x)\chi(x),\;\;\;
    P^\alpha(x)  = \bar{\chi}(x)\gamma_5\frac{\tau^\alpha}{2}\chi(x),  \\
    A_\mu^\alpha(x) & =
    \bar{\chi}(x)\gamma_\mu\gamma_5\frac{\tau^\alpha}{2}\chi(x),
    \;\;\;
    V_\mu^\alpha(x) = \bar{\chi}(x)\gamma_\mu\frac{\tau^\alpha}{2}\chi(x), \\
    T_{\mu\nu}^\alpha(x) & = \bar{\chi}(x)\sigma_{\mu\nu}\frac{\tau^a}{2}\chi(x), 
    \;\;\;
    T_{\mu\nu}^0(x) = \bar{\chi}(x)\sigma_{\mu\nu}\chi(x) 
    \; .
    \label{operators}
  \end{split}
\end{equation}
and ii) the twist angle $\omega$, where 
$\tan \omega = \mu_q/(m_0 - m_\mathrm{crit})$ and $am_\mathrm{crit}$
is determined as discussed in section~\ref{sec:maximaltwist}.
The expressions we get are
\begin{equation}
  \label{eq:1}
  A'^{\alpha}_\mu =
  \begin{cases}
    \cos(\omega)A_\mu^\alpha+\epsilon^{3\alpha\beta}\sin(\omega)
V_\mu^\beta & \text{($\alpha=1,2$)},\\
    A_\mu^3 & \text{($\alpha=3$)},
  \end{cases}
\end{equation}
\begin{equation}
  \label{eq:2}
  V'^{\alpha}_\mu =
  \begin{cases}
    \cos(\omega)V_\mu^\alpha+\epsilon^{3\alpha\beta}\sin(\omega)
A_\mu^\beta & \text{($\alpha=1,2$)},\\
    V_\mu^3 & \text{($\alpha=3$)},
  \end{cases}
\end{equation}
\begin{equation}
  \label{eq:3}
  P'^{\alpha} =
  \begin{cases}
    \cos(\omega)P^3+i\frac{1}{2}\sin(\omega) S^0 & \text{($\alpha=3$)},\\
    P_\alpha & \text{($\alpha=1,2$)},
  \end{cases}
\end{equation}
  \label{eq:4}
\begin{equation}
  S'^{0} =
    \cos(\omega)S^0+2i\sin(\omega) P^3\; ,
\end{equation}
\begin{equation}
  \label{eq:5}
  T'^{\alpha}_{\mu\nu} =
  \begin{cases}
    T^{\alpha}_{\mu\nu} & \text{($\alpha=1,2$)}, \\
    \cos(\omega)T_{\mu\nu}^3 -i \frac{1}{2}\epsilon^{\mu\nu\lambda\rho}\sin(\omega)
T_{\lambda\rho}^0 & \text{($\alpha=3$)}\; .
  \end{cases}
\end{equation}
These expressions follow from  
the relation between twisted basis ($\chi$) and
physical basis ($\psi$) quark fields, which 
(see eq.(\ref{eq:Sf}) and ref.~\cite{Frezzotti:2003ni}) reads 
\begin{equation}
\chi \; = \; e^{-i\gamma_5\tau^3\omega/2} \psi \; , \qquad 
\bar\chi \; = \; \bar\psi e^{-i\gamma_5\tau^3\omega/2} \; ,
\label{eq:twi2phy}
\end{equation}
and the (obvious) definitions of
the bare primed operators in terms
of physical basis quark fields ($\alpha=1,2,3$) 
\begin{equation}
  \begin{split}
  A'^\alpha_\mu & = 
\bar{\psi}(x)\gamma_\mu\gamma_5\frac{\tau^\alpha}{2}\psi(x),
\qquad     
  V'^\alpha_\mu = 
\bar{\psi}(x)\gamma_\mu\frac{\tau^\alpha}{2}\psi(x),
 \\
  P'^\alpha & = 
\bar{\psi}(x)\gamma_5\frac{\tau^\alpha}{2}\psi(x),
\qquad
 T'^\alpha_{\mu\nu}(x) = \bar{\psi}(x)\sigma_{\mu\mu}\frac{\tau^a}{2}\psi(x),
\\
 S'^0 & = \bar\psi(x) \psi(x)  
 \; .
  \end{split}
\end{equation}
All these bare operators renormalize multiplicatively, with the exceptions of
$P'^3$ and $S'^0$, which undergo an additive mixing with the identity (cubically
divergent for $P'^3$, quadratically divergent and vanishing as $\mu_q \to 0$
for $S'^0$).
For the expression of renormalization constants as functions of $\omega$ and
the renormalization constants of  standard Wilson quark bilinears and
further details, see ref.~\cite{Frezzotti:2003ni}. It should be remarked that
substantial simplifications occur for $\omega = \pm \pi/2$
(maximal twist) in almost all formulae above. Moreover at maximal twist
also the formulae for renormalization constants~\cite{Frezzotti:2003ni} 
get much simpler than at generic $\omega$.

\section{Evaluation of disconnected loops}
\label{sec:appendixB}

The quark-disconnected (simply ``disconnected'' in the following for
brevity) components of correlators are intrinsically noisier than the
connected components, so it is essential to evaluate them as accurately
as possible. For this purpose we need to compute the disconnected loops
at every $t$ value and for as many gauge configurations as possible.
This can  be achieved by using the stochastic source methods as we now
discuss.  The goal of the approach is to have an error arising from the
stochastic nature of the method  which is smaller than the intrinsic
variability  associated with varying $t$ and gauge configuration. If
this is achieved, then  the stochastic error is negligible in the sense
that any further improvement in the  signal can only be obtained if more
gauge configurations are employed.

As discussed in sect.~\ref{subsec:sources}, the basic idea is to use stochastic 
sources ($\xi$) having in general support on the whole lattice and solve the linear system 
for the quantities 
 \begin{equation}
 \phi = M^{-1} \xi  \, ,
\label{eq:phi-1} 
 \end{equation}
where $M$ is the lattice Dirac matrix for a given flavour. The equation
above is the same as eq.~(\ref{oneend2}), with the omission of the noise
sample label $r$ (to lighten notation). Note also that in this appendix
the normalization of $M$ is taken such that, if $D_{\rm latt}$ 
denotes the two-flavour Dirac matrix in eq.~(\ref{eq:Sf}), then 
 \begin{equation}
M_u = 2\kappa {\rm tr}[ aD_{\rm latt} (1+\tau_3)/2 ] = A + H \, ,
\qquad A = 1+ 2\kappa a\mu_q i\gamma_5 \, ,
\label{eq:Dirnota}
 \end{equation}
with $H$ the usual Wilson first-neighbour hopping matrix.
It follows that
 \begin{equation}
 \sum [\xi^* X \phi ]_R =
 \sum X M^{-1} + {\rm noise}
 \label{eq:B1}
 \end{equation}
 where the symbol $[...]_R$ refers (as in sect.~\ref{subsec:sources}) to the average over $R$ 
 samples of the stochastic source, the symbol $\sum$ denotes the sum
 over colour, spin and space-time indices and $X$ can be (almost) any structure we
 wish to evaluate, like $\gamma$-matrix, gauge links, Fourier factor, $\cos(kx)$,
 etc... It should be observed that in evaluating the disconnected
contributions to the neutral  meson correlators each one of the two
quark loops arising from Wick contractions 
 must be averaged over completely {\em independent} samples of
stochastic sources for the purpose of avoiding unwanted biases. 
 Moreover, for each quark loop diagram, the sum in eq.~(\ref{eq:B1}) is  
restricted to one single time-slice, while still ranging over all color,
spin and space indices.

A method we employed to reduce the variance of the stochastic noise without much
additional computational effort is the hopping-parameter
method~\cite{McNeile:2000xx}. This relies on the observations that the
 first four terms in the  hopping parameter expansion of $\sum  X M^{-1}$
can be easily evaluated exactly on each gauge configuration and that replacing 
their stochastic estimates with the exact values significantly reduces the variance. 
In fact, writing $M_u$ (see eq.~(\ref{eq:Dirnota})) in the form $M_u=(1+HB)A$,
where $B=1/A$, one easily obtains the identity 
 \begin{equation}
 1/M_u=B-BHB+B (HB)^2 - B (HB)^3 + (1/M_u) (HB)^4\, ,
 \label{eq:B2}
 \end{equation}
 which can be used  to give
 \begin{equation}
 \sum X/M_u \; = \; \sum \left\{   X (B-BHB+B (HB)^2 -B (HB)^3 + (1/M_u) (HB)^4  ) \right\} \, .
 \label{eq:B3}
 \end{equation}
 The last term in eq.~(\ref{eq:B3}) can be evaluated stochastically because 
 \begin{equation}
   X (1/M_u) (HB)^4   = \lim_{R\to \infty}  [ \xi^*  (HB)^4 X \phi ]_R
 \label{eq:B4}
 \end{equation}
 Since $H^{\dag}=\gamma_5 H \gamma_5$ and
$\gamma_5$ commutes with $B$, the last formula can be rewritten in the form 
 \begin{equation}
  X (1/M_u) (HB)^4   = \lim_{R\to \infty} [ (\gamma_5 (B^\dagger H)^4 \gamma_5\xi)^* X \phi ]_R \, .
 \label{eq:B5}
 \end{equation}
 Thus four extra multiplications of the source $\xi$ by $B^\dagger H$ are
 needed. This is a negligible overhead compared to the inversion needed to
 obtain $\phi$.  The first four terms in eq.~(\ref{eq:B3})
 do not involve $1/M_u$ and can be, as said above, evaluated straightforwardly
 for any choice of $X$. 
 For a local operator $X$, the only non-zero contributions are from
 the first term if $X$ is proportional to $1$ or $\gamma_5$ and the third
 term  if $X$ is proportional to
 $\gamma_5$. 
 For a non-local operator $X$ whose length of spatial path is more than four
 lattice hops (as  used in this paper), the first four terms are all zero. 

This variance reduction method reduces the standard error of the stochastic
samples by a factor of 1.5 or more in our case. This is valuable (it saves a
factor 2-3 in computational time), but for twisted mass QCD a much more powerful
method is also available, although it applies only to the case 
$\Sigma  X (1/M_u - 1/M_d)$. This last  method can be, and has been indeed, used in 
many important applications, essentially all those where one has to evaluate correlators 
with insertions of neutral meson operators of the form (in the twisted basis) $\bar\chi
\Gamma \tau_3 \chi$, with any Dirac matrix $\Gamma$ and the flavour matrix $\tau_3$. 
Interpolating fields of this type occur e.g.\ in the two-point correlators for $\eta'$, $f_0$ 
and $\pi^0$ mesons (actually only one of the possible operators for $\pi^0$), 
as one can see from the 
table for the (twisted basis) neutral meson operators reported in sect.~\ref{sec:neutralmesons}.

The powerful method alluded above relies on combining the identities
 \begin{equation}
(M_d-M_u) =-4i  \kappa  a \mu_q  \gamma_5
 \label{eq:B6}
 \end{equation}
and
 \begin{equation}
   (1/M_d) (M_d-M_u) (1/M_u) = 1/M_u -1/M_d 
 \label{eq:B7}
 \end{equation}
to get
 \begin{equation}
  1/M_u - 1/M_d =  -4i \kappa a\mu_q (1/M_d) \gamma_5  (1/M_u) \, .
 \label{eq:B8}
 \end{equation}
 The latter relation already serves as a method of variance reduction
because the explicit (small, in our simulations) factor of $a\mu_q$
reduces the magnitude of the fluctuations. On top of that,  an even more
important point is that the r.h.s.\ of eq.~(\ref{eq:B8}) can be
evaluated very effectively with the help of the
``one-end-trick''~\cite{Foster:1998vw,McNeile:2006bz} and no further
inversions. In fact, since $M_u^{\dag}=\gamma_5 M_d \gamma_5$, one has 
 \begin{equation}
  \sum X (1/M_u - 1/M_d) =
    -4i \kappa a\mu_q \sum X \gamma_5 (1/M_u)^{\dag}   (1/M_u) \, ,
 \label{eq:B9}
 \end{equation}
which can be evaluated with noise/signal ratio of O(1) via
 \begin{equation}
  \sum X (1/M_u - 1/M_d) =
    -4i \kappa a\mu_q \sum [  \phi^* X \gamma_5 \phi ]_R + {\rm noise} \, ,
 \label{eq:B10}
 \end{equation}
 where (we recall) $ \phi=(1/M_u) \xi$ and   $\phi^* = \xi^*
(1/M_u)^{\dag} $. Apart from the explicit sum denoted by $\sum$, the
r.h.s.\ of this formula contains an implicit sum over the space-time
indices of the stochastic source $\xi$ in $\phi$  and $\phi^*$, which
contributes to reduce the variance as it receives contributions from the
whole lattice (space-time) volume.

To give an idea of the effectiveness of the method based on eq.~(\ref{eq:B10})
we consider, as an example, the special case $X=i \gamma_5$, where one obtains
 \begin{equation}
\sum i \gamma_5 (1/M_u - 1/M_d) =
    4 \kappa a\mu_q \sum [ \phi^*  \phi ]_R + {\rm noise} \, .
 \label{eq:B11}
 \end{equation} 
 At $\beta=3.9$ and $\mu_q=0.004$ (ensemble $B_1$) 
the method based on eq.~(\ref{eq:B11}) yields an error 
which turns out to be 6 times smaller than what would be obtained 
with a conventional stochastic volume source.
 From the measured stochastic contribution to the signal, as well 
as the observed total fluctuation, one can extract the intrinsic
variation stemming from the statistical fluctuations of the gauge field.
The goal of the stochastic method is to have errors arising from the 
stochastic method which are negligible compared to the intrinsic (gauge) noise.
 This we achieve,  finding that the stochastic contribution to  the
total error has a standard deviation which is 2/3 of the standard
deviation  arising from the intrinsic variation of the signal. In the
example, above we employed 24 stochastic sources (with no components set
to zero), resulting in a cost of 24 inversions, per gauge configuration.
Note that a similar number of inversions is needed to compute the
(quark-connected) charged  meson correlators. 

We thus find that this variance reduction method, where applicable, is
very powerful and effectively reduces the stochastic noise in the
neutral meson correlators, making it smaller than the intrinsic noise 
coming from the fluctuations of the gauge field.

\section{$\Gamma$-method and data-blocking}
\label{sec:appendixC}

In this appendix, we discuss the $\Gamma$-method and the data-blocking procedure
we have used to estimate the statistical errors of our physical observables. 

\subsection{$\Gamma$-method}
\label{sec:newC_1}

In this section, for completeness, we just recall the basis of the $\Gamma$-method introduced in~\cite{Wolff:2003sm}.
In the case of a {\em primary} stochastic variable with ``true value'' $A$
(the symbol $A$ will also be used to denote the observable itself), 
a suitable estimator of the error on the ensemble average $\bar a$, i.e.\ its standard deviation
$\sigma_{\bar a}$, is given by~\footnote{For a discussion of these issues, see~\cite{Sokal:1989ea}
and references therein.}
\begin{equation}\label{eq:est_sigma}
\sigma^2_{\bar a}=\frac{1}{N}\sum_{n=-W}^{W}\Gamma_{\bar a}(n)\, ,
\end{equation}
where $N$ is the number of measurements, $2W+1 \ll N$ is the number of consecutive measurements 
used in the estimation (measurement ``window'') and
\begin{equation}\label{eq:est_gamma}
\Gamma_{\bar a}(n) = \frac{1}{N-|n|} \sum_{i=1}^{N-|n|}(a^i-\bar a) (a^{i+|n|}- \bar a) \, .
\end{equation}
Here $\Gamma_{\bar a}(n)$ represents the straightforward estimator of the autocorrelation function
$\Gamma_A(n)=\langle (a^i-A) (a^{i+|n|}-A)\rangle$ (the index $i$ in $a^i$ labels the individual measurements,
while $\langle \dots \rangle$ denotes the theoretical expectation value).

The integrated autocorrelation time is conventionally
defined for primary quantities as in eq.~(\ref{eq:tauint})
and estimated by (see eq.~(\ref{eq:est_sigma}))
\begin{equation}\label{eq:est_tau_int}
  \tau_\mathrm{int}({\bar a})=\frac{1}{2\Gamma_{\bar a}(0)}\sum_{n=-W}^{W}\Gamma_{\bar a}(n)\equiv 
  \frac{N \sigma^2_{\bar a}}{2\bar\sigma^2_{a}}
  \, ,
\end{equation}
Note that $\Gamma_{\bar a}(0)\equiv\bar\sigma^2_{a}$, see eq.~(\ref{eq:est_gamma}), is
an estimate of the {\it a priori} variance of $A$.

The $\Gamma$-method can also be applied to the analysis of secondary observables, $F= f(A)$, 
where $f$ denotes a non-linear function of several primary observables, $A \equiv \{A_1, A_2, \dots\}$.
A typical example is the case where $A$ is given by the values of two-point hadron correlators
at different time separations, with different smearing levels, {\em etc.}, while $F$
is a suitable estimator of the hadron mass; of course, the details of the function $F=f(A)$ 
depend on the specific choice of the estimator, e.g.\ on the form of the fit ansatz 
for the  correlators and the range of time separations employed in the fit.   
The main point here is that the deviation of 
any given finite-statistics estimate of $F$, $\bar F \equiv f(\bar a)$,  
from the true value $f(A)$ can 
be approximated, in the limit of large statistics, by retaining the  
first term of the Taylor expansion of $f(\bar a)$ around $f(A)$, i.e.\ by writing 
\begin{equation}\label{eq:taylor}
f(\bar a)-f(A)\simeq \sum_\alpha\frac{\partial f(A)}{\partial A_\alpha}(\bar a_\alpha-A_\alpha)\, ,
\end{equation}
where $\alpha$ is the index labeling the primary quantities, $A_\alpha$, upon which $f$ depends.
This remark suggests to define a new quantity, $A_f$, which is a simple linear combination of primary
quantities, and the corresponding finite-statistics estimate, $\bar{a}_f$, via the formula 
\begin{equation}\label{eq:lin_var}
A_f\equiv\sum_\alpha\frac{\partial f(A)}{\partial A_\alpha} A_\alpha\, ,\quad\quad
\bar{a}_f\equiv\sum_\alpha\frac{\partial f(A)}{\partial A_\alpha} \bar{a}_\alpha\, ,
\end{equation}
where $\bar{a}_\alpha$ is the ensemble average of the primary stochastic variable $a_\alpha$
(with ``true value'' $A_\alpha$, as above). The variance of $\bar F =f(\bar a)$ 
will be given by 
\begin{equation}
\label{eq:th_var_F}
\sigma^2_{\bar F}\equiv\langle(f(\bar a)-f(A))^2\rangle\simeq\langle(\bar a_f-A_f)^2\rangle\, ,
\end{equation}
where the truncation of the Taylor series produces a relative bias O($N^{-1}$) which can be
neglected if the number of measurements $N$ is sufficiently large. A further bias of the same 
order of magnitude arises from the replacement $\frac{\partial f(A)}{\partial A_\alpha}
\rightarrow  \left.\frac{\partial f(A)}{\partial A_\alpha}\right|_{A = \bar a}$ in eq.~(\ref{eq:lin_var}),
which is done in practice to evaluate the first derivatives of $f$ with respect to the $A_\alpha$'s.  
At this point $\sigma^2_{\bar F}$ is estimated by the formula that is obtained from 
eq.~(\ref{eq:est_sigma}) by replacing $\Gamma_{\bar a}(n)$ with
\begin{equation}\label{eq:est_gamma_F}
\Gamma_{\bar a_f}(n) = \frac{1}{N-|n|} \sum_{i=1}^{N-|n|}(a_f^i-\bar a_f) (a_f^{i+|n|}- \bar a_f) \, .
\end{equation}

\subsection{Binning method}
\label{sec:newC_2}

In the case where a data-blocking (also called binning) procedure 
is instead adopted to account for autocorrelations,
the bin-size $B$ plays a role similar to that of the window $W$
in the $\Gamma$-method.  The integrated autocorrelation time can thus 
be estimated, for sufficiently large values of $B$, by 
\begin{equation}
  \tau_\mathrm{int}({\bar F})\simeq
   \frac{\sigma^2_{\bar F}(B)}{2\sigma^2_{\bar F}(1)} \, ,
\end{equation}
where $\sigma_{\bar F}(B)$ denotes the jackknife estimate
of the error on $\bar F$ (the mean value of $F$) that is obtained upon binning
the measurements into blocks of size $B$. 

\subsection{Error on the error: $\Gamma$-method {\it vs} data-blocking}
\label{sec:new_C3}

The estimator of eq.~(\ref{eq:est_sigma}) 
allows to reach the optimal compromise between the relative statistical
error on $\bar\sigma_{\bar a}$ raising with $\sqrt{W}$,
i.e.\ $\delta_\mathrm{stat}(\bar\sigma_{\bar a})/\bar\sigma_{\bar a}\sim \sqrt{W/N}$, and the
relative systematic error (bias) decreasing exponentially with $W$,
i.e.\ $\delta_\mathrm{syst}(\bar\sigma_{\bar a})\sim 1/2\exp{(-W/\tau)}$, where $\tau$ is the
characteristic time of slowest exponential mode
of $\Gamma(n)$ (exponential autocorrelation time). An ``optimal'' value, $W_\mathrm{opt}$,
to be used as upper and lower bound for the sum in eq.~(\ref{eq:est_sigma})
can be obtained, e.g.\ by gradually increasing $W$ and
inspecting ``by eye'' the onset of a plateau for $\bar\sigma_{\bar a}$
as a function of $W$, or requiring minimisation of the total error
$\delta_\mathrm{tot}=\delta_\mathrm{stat}+\delta_\mathrm{syst}$~\cite{Wolff:2003sm}.
Any valid criterion to truncate the sum necessarily corresponds to values of $W_{opt}$
for which the truncation errors become comparable with the statistical
noise level on $\bar\sigma_{\bar a}$. This choice corresponds to an uncertainty on the
error on $\bar\sigma_{\bar a}$ decreasing like $\sim {\mathrm O}(N^{-1/2})$.
For comparison we recall that the error on $\bar\sigma_{\bar a}$ upon use of
the binning method would decrease only like $\sim {\mathrm O}(N^{-1/3})$~\cite{Wolff:2003sm}.
In this case in fact the optimal choice corresponds to find a compromise 
between the relative statistical error on $\bar\sigma_{\bar a}$ 
(i.e.\ $\delta_\mathrm{stat}(\bar\sigma_{\bar a})/\bar\sigma_{\bar a}\sim \sqrt{B/2N}$) 
which increases with $\sqrt{B}$, and the relative systematic error (bias) 
(i.e.\ $\delta_\mathrm{syst}(\bar\sigma_{\bar a})\sim \tau/2B$) which decreases with $B^{-1}$.

\subsection{Further remarks}
\label{sec:newC_4}

In our error analysis carried out using the $\Gamma$-method,
we decided to compare different criteria for the windowing procedure
in order to test in this respect the robustness of our estimates. 
One method is given by the algorithm proposed in~\cite{Wolff:2003sm}
which is close to optimal. A second criterion, which is slightly more conservative, consists in stopping
the procedure as soon as $\bar\Gamma(n)$ becomes negative due to statistical fluctuations.
In the 15 analysed cases (5 simulation points times 3 quantities), no systematic trend
could be detected, with the two methods giving in most of the cases similar results.
In the cases where we cyclically vary the time the wall source over the lattice 
(see sect.~\ref{subsec:sources})
in order to restore translation invariance in the MC time, as required by the $\Gamma$-method,
we average beforehand 
correlators over source cycles~\footnote{We generically find correlations 
between consecutive measurements taken on well 
separated time-slices (e.g.\ by $\Delta t=12a$) to be negligible.}.
We recall that the time-slice sequences used for the different ensembles and the value
of $n=t_p$ are specified in Table~\ref{tab:method}. 

As already mentioned, the results of the $\Gamma$-method have been
checked against binning procedures. 
For observables that are non-linear functions of the primary quantities,
the error estimates were  
obtained by combining the binning
procedure with either bootstrap-sampling 
(with bin sizes 
$B=4,8,16,32$ in trajectory units)
or standard jackknife. 
In the latter case the optimal bin size
$B_{opt}$ was determined by requiring stabilization of the estimate of
the error (with $B_{opt}/\tau_{int}\approx 10$ or larger).

Different methods give in general comparable results. In the case of
the binning+bootstrap procedure stabilization of the error is however not always evident
at the maximal bin size ($32$ in trajectory units). 
In particular the PCAC quark mass turns out to be affected by significant
autocorrelations (see sect.~\ref{subsec:autocorr})
and the binning procedure seems not to be able to give reliable estimates of the error.
In this case indeed the results lie systematically below the estimates from
the $\Gamma$-method. This can be understood recalling that the $\Gamma$-method 
leads to a more favourable dependence upon the number of measurements in the 
error attributed to the autocorrelation time than the binning method.

In view of these findings we have decided to use the $\Gamma$-method for the 
estimates of the errors on the plaquette and $am_{\rm PCAC}$.
Also the error estimates for fermionic quantities (other than $am_{\rm PCAC}$)
quoted in sect.~\ref{sec:results} come from this method. However, 
similar results are obtained if a binning based procedure is employed.

\section{Details of the static potential calculation}
\label{sec:appendixD}

In this appendix we provide some details on the way we
compute the static quark-antiquark potential
from our dynamical gauge configurations.

\subsection{Improved static action}

An improvement on the signal-to-noise ratio in the measurements
of the Wilson loop can be obtained by
employing suitably smeared temporal links. This can be viewed as a
convenient modification (or improvement) of the action for static
quarks~\cite{DellaMorte:2003mn,DellaMorte:2005yc}
as long as gauge invariance, cubic and
parity symmetries as well as the local conservation of the
static quark number and the static quark spin symmetry are preserved.
Under these conditions
it is still guaranteed that the static quark action is free from
O($a$) cutoff effects~\cite{Kurth:2000ki}.
The statistical improvement alluded above comes from a reduction of the
noise-to-signal ratio
essentially stemming from the fact that the modified static quark action
obtained via the use
of smeared temporal links induces a self-energy mass term with a
significantly reduced coefficient
in front of the $a^{-1}$ term~\cite{DellaMorte:2005yc}.
For our measurements we use the so-called HYP-improved static quark action,
which is obtained by replacing the temporal links $U_4(\vec x,x_0)$
in the Wilson loop by HYP-smeared links~\cite{hasenfratz:2001hp}
\begin{equation}
U_4(\vec x, x_0) \rightarrow V^\text{HYP}_4(\vec x,x_0) \, .
\end{equation}
The HYP-smearing requires the specification of three
parameters $\vec \alpha = (\alpha_1, \alpha_2, \alpha_3)$
and, following ref.~\cite{DellaMorte:2005yc}, we choose
$\vec \alpha = (1.0,1.0,0.5)$ throughout our calculation.

\subsection{Spatial smearing}
\label{subsec:appendixDfuzzing}

The smoothing of the spatial links has the effect of
reducing excited-state contamination in the correlation functions of the
Wilson loops in the potential measurements. The operators which we
measured in
the simulations are constructed using the spatial APE smearing of
ref.~\cite{Albanese:1987ds}.
The smoothing procedure we use consists in replacing every spatial link
$U_j(x), j=1,2,3$
by itself plus a sum of its neighbouring spatial staples and then
projecting back to the nearest
element in the SU(3) group, {\em i.e.}\ we write 
\begin{eqnarray}
  {\cal S}_1 U_j(x) & \equiv & {\cal P}_{\text{SU(3)}} \Big\{ U_j(x) +
\lambda_s \sum_{k \neq j}
  (U_k(x) U_j(x+\hat k) U_k^\dagger(x+\hat j)  \\
 & &  \hspace{2cm}  + U_k^\dagger(x-\hat k)
  U_j(x-\hat k) U_k(x-\hat k+\hat j))  \Big\}\, . \nonumber
\end{eqnarray}
Here, ${\cal P}_{\text{SU(3)}} Q$ denotes the unique projection onto the
SU(3) group element $W$, which maximises $\text{Re} \text{Tr}(W Q^\dagger)$ for any
$3 \times 3$ matrix $Q$. The smeared and SU(3) projected link ${\cal S}_1
U_j(x)$ retains all the symmetry properties of the original link $U_j(x)$
under gauge transformations, charge conjugation, reflections and permutations
of the coordinate axes.  The whole set of spatially smeared links,
$\{{\cal S}_1 U_j(x), x \epsilon L^4\}$, forms the spatially smeared gauge
field configuration. An operator $\cal O$ which is measured on a $n$-times
iteratively smeared gauge field configuration is called an operator at
smearing level ${\cal S}_n$, indicated by the symbol ${\cal S}_n
{\cal O}$. From our experience a good choice is to use $M=5$ different smearing
levels ${\cal S}_n$, with $n=8,16,24,32,40$, and in all cases a smearing parameter
$\lambda_s=0.25$.

\subsection{Static quark-antiquark pair correlators}

The matrix of static quark-antiquark pair correlation functions, each of
which from a technical viewpoint corresponds to a spatially smeared and
temporally improved Wilson loop, is constructed in the following way.
At fixed $x_0$ we first form smeared string (i.e.\ quark-antiquark pair)
operators along the three spatial axes, connecting $\vec x$ with $\vec x + r \hat
i$, given by 
\begin{multline}
  \label{eq:smearedspatiallink}
  {\cal S}_n V_i(\vec x, \vec x + r \hat i;x_0) = \\
  {\cal S}_n U_i(\vec x,x_0) {\cal S}_n U_i(\vec x+a\hat i,x_0) \ldots
{\cal S}_n U_i(\vec x + (r-a)
  \hat i,x_0), \quad i=1,2,3\, ,
\end{multline}
and improved temporal links at fixed $\vec x$, connecting $x_0$ with
$x_0+t$, given by 
\begin{equation}
  \label{eq:unsmearedtemporallink}
  V_4(x_0,x_0+t;\vec x) = V^\text{HYP}_4(\vec x,x_0) V^\text{HYP}_4(\vec
x,x_0+a)
\ldots V^\text{HYP}_4(\vec x,x_0+(t-a)) \, .
\end{equation}
The smeared Wilson loop~\footnote{Let us remark that we measure the on-axis
potential only, {\em i.e.}\ the potential extracted from Wilson loops
having spatial extent in the direction of the lattice axes $\hat i, i=1,2,3$
only.} is then obtained by computing
\begin{multline}
  \label{eq:smearedWilsonloop}
  {\cal W}_{lm}(r,t) = \sum_{\vec x, x_0} \sum_{i=1}^3 \text{Tr} \, {\cal
S}_l
  V_i(\vec x, \vec x + r \hat i;x_0) V_4(x_0,x_0+t;\vec x + r \hat i) \\
{\cal S}_m
  V_i^\dagger(\vec x, \vec x + r \hat i;x_0+t) V_4^\dagger(x_0,x_0+t;\vec
x) \, .
\end{multline}
Finally we define the matrix of static quark-antiquark pair correlators
according to the formula
\begin{equation}
  \label{eq:Wilsonloopcorrelation}
  C_{lm}(r,t) = \langle {\cal W}_{lm}(r,t) \rangle = C_{ml}(r,t)\, ,
\end{equation}
 where the average is over the configurations of the ensemble.
Since we have chosen to employ $M=5$ different string operators (as
discussed above) and
we are concerned with correlators where two such operators are
inserted, we
end up with a $5 \times 5$ matrix of static quark-antiquark pair correlators.

\bibliographystyle{h-physrev4} 
\bibliography{paper} 
 
\end{document}